\documentclass[authoryear]{elsarticle}
\usepackage{graphicx}        	
\usepackage{amsmath,amssymb,amsthm}
\usepackage[utf8]{inputenc}
\usepackage[T1]{fontenc}
\usepackage{bm}
\usepackage{subfig}
\usepackage{url}
\usepackage{ulem}
\usepackage{hyperref}
\usepackage[dvipsnames]{xcolor}
\hypersetup{
    colorlinks,
    linkcolor={red!50!black},
    citecolor={blue!80!black},
    urlcolor={blue!80!black}
}
\usepackage{rotating}

\newcommand{\vecr}{\mathbf{r}}

\newcommand{\vecq}{\mathbf{q}}

\newcommand{\vecw}{\mathbf{c}}

\newcommand{\vecu}{\hat{\mathbf{q}}}

\newcommand{\vecx}{\mathbf{x}}
\newcommand{\vecy}{\mathbf{y}}

\newcommand{\noise}{W}

\DeclareMathOperator{\Tr}{Tr}

\theoremstyle{definition}

\begin{document}

\title{Bayesian uncertainty quantification in linear models for diffusion MRI}
%

\author[elekta,liu,cmiv]{Jens Sj\"olund\corref{cor1}}
\ead{jens.sjolund@liu.se}

\author[liu,cmiv,ida]{Anders Eklund}
\author[liu,cmiv]{Evren Özarslan}
\author[math]{Magnus Herberthson}
\author[SICS]{Maria Bånkestad}
\author[liu,cmiv]{Hans Knutsson}

\cortext[cor1]{Corresponding author}
\address[elekta]{Elekta Instrument AB, Kungstensgatan 18, Box 7593, SE-103 93 Stockholm, Sweden}
\address[liu]{Department of Biomedical Engineering, Link\"oping University, Link\"oping, Sweden}
\address[cmiv]{Center for Medical Image Science and Visualization (CMIV), Link\"oping University, Sweden}
\address[ida]{Department of Computer and Information Science, Link\"{o}ping University, Link\"{o}ping, Sweden}
\address[math]{Department of Mathematics, Link\"{o}ping University, Link\"{o}ping, Sweden}
\address[SICS]{RISE SICS, Isafjordsgatan 22, Box 1263, SE-164 29 Kista, Sweden}

\date{\today}

\begin{abstract}
Diffusion MRI (dMRI) is a valuable tool in the assessment of tissue microstructure. By fitting a model to the dMRI signal it is possible to derive various quantitative features. Several of the most popular dMRI signal models are expansions in an appropriately chosen basis, where the coefficients are determined using some variation of least-squares. However, such approaches lack any notion of uncertainty, which could be valuable in e.g. group analyses.
In this work, we use a probabilistic interpretation of linear least-squares methods to recast popular dMRI models as Bayesian ones.  This makes it possible to quantify the uncertainty of any derived quantity. In particular, for quantities that are affine functions of the coefficients, the posterior distribution can be expressed in closed-form. 
We simulated measurements from single- and double-tensor models where the correct values of several quantities are known, to validate that the theoretically derived quantiles agree with those observed empirically. We included results from residual bootstrap for comparison and found good agreement.
The validation employed several different models: Diffusion Tensor Imaging (DTI), Mean Apparent Propagator MRI (MAP-MRI) and Constrained Spherical Deconvolution (CSD). We also used in vivo data to visualize maps of quantitative features and corresponding uncertainties, and to show how our approach can be used in a group analysis to downweight subjects with high uncertainty.
In summary, we convert successful linear models for dMRI signal estimation to probabilistic models, capable of accurate uncertainty quantification.
\end{abstract}

\maketitle


\section{Introduction}
\label{sec:intro}

Diffusion magnetic resonance imaging (dMRI) permits the noninvasive assessment of tissue microstructure. By fitting a model of the dMRI signal in each voxel it is possible to
derive various quantitative features. Examples include voxel-based scalar indices such as fractional anisotropy (FA) or return to origin probability (RTOP), as well as the pairwise probability of a virtual fiber being traced between two points. Using such measures it is possible to perform statistical group analyses. For the reliability of such tests it is essential to quantify the uncertainty of the relevant measures. This fact is well-established in the field of functional magnetic resonance imaging (fMRI) --- where it is common to, for example, downweight subjects with a high variance \citep{Chen2012, Woolrich2004} --- but not so for dMRI; the most popular approach for doing FA group comparisons \citep{Smith2006} ignores uncertainty in FA. 

A large body of research has been devoted to formulating new diffusion models that address the inability of diffusion tensor imaging (DTI) \citep{Basser1994} to resolve crossing and kissing fibers. A common trait among several of the most widely used methods is that they expand the signal in an appropriately chosen functional basis. Often, the coefficients of the expansion are then determined by some variation of linear least-squares (see Table \ref{tab:linearModels} for a non-exhaustive list). This is precisely the type of models we are concerned with in this paper --- linear (in the coefficients) models fitted with least-squares. Incidentally, some non-parametric models \citep{Andersson2015, Sjolund2017} also belong to this class, although we will not elaborate on the connection here. Our key observation is that linear models fitted with least-squares are amenable to a probabilistic reinterpretation. We will show that it follows almost immediately that, under the same assumptions as in the fitting, it is possible to determine the full posterior distribution of the coefficients --- not just a point estimate.

This is, of course, not the first time someone has taken a probabilistic view on signal estimation in dMRI. It is well-known that the noisy signal in MRI follows a Rician, or more generally a non-central Chi, distribution \citep{Gudbjartsson1995}, which on the other hand is approximately Gaussian when the signal-to-noise ratio is, at least, moderately high ($\gtrsim 3$). In DTI \citep{Basser1994} and diffusion kurtosis imaging (DKI) \citep{Jensen2005}, it is common to fit a linear model to the logarithm of the signal. The resulting log-Rician distribution is again approximately normal for moderate signal-to-noise ratios, but with signal-dependent (heteroscedastic) noise \citep{Salvador2005}. In such cases, weighted least-squares has been shown to work well \citep{Rawlings2001, Veraart2013}. 
Error propagation in DTI fitted with nonlinear least-squares has also been investigated \citep{Koay2007}.

Bayesian methods \citep{Behrens2003, Gelman2013, Gu2017, Wegmann2017} are distinctly different from methods using least-squares to find point estimates. By assuming parametric probability distributions for the likelihood and for every parameter in the model, Bayesian methods make it possible to derive the probability distribution of any quantity of interest, at least in principle. Actually evaluating such distributions, however, typically relies extensively on sampling methods such as Markov Chain Monte Carlo.

Bootstrapping is a frequentistic alternative to Bayesian models. The general idea is to approximate the underlying probability distribution with an empirical one. Samples are drawn with replacement and for each draw the parameter of interest is calculated. By repeating this procedure, it is possible to approximate the sampling distribution of the relevant parameter. Naturally, the quality of the approximation degrades as the number of empirical samples decreases. 
Residual bootstrap \citep{Chung2006} and wild bootstrap \citep{Whitcher2008} are two forms of model-based bootstrapping that have been applied to dMRI. In residual bootstrap, the normalized residuals (after fitting a model) are resampled, and the fitting procedure is repeated for this new draw, after which the parameter value of interest is recorded. All residuals are assumed to have identical distributions and resampling is done freely among them. In wild bootstrap, on the other hand, modified residuals are randomly added or subtracted to the fitted point where they originated from, without being distributed to other design points.
Model-based bootstrapping is only reliable insofar as the model can adequately describe the measured diffusion signals \citep{Yuan2008}.  
Basically all of the methods described above have been applied to tractography \citep{Behrens2007, Berman2008, Haroon2009, Jeurissen2011, Jones2003, Jones2008}.

In this work, instead of starting out by assuming more or less contrived prior distributions for the coefficients and the likelihood, we look at methods that are tried and tested and see what the corresponding priors are. Surprisingly, it results in a simple closed form expression for the posterior distribution of coefficients, and by extension also of all parameters linear in the coefficients. In other cases
there is no need to repeat the whole fitting procedure, as in bootstrapping methods. Since the posterior is available in closed form it is very efficient to sample from it directly.

\begin{table}
\centering
\makebox[\textwidth][c]{
\begin{tabular}{lccccc}
Method & $y$ & $\phi(\vecx)$ & $\vecw$ & $W$ & $\Lambda$\\[2pt]
\hline
\noalign{\vskip 2pt}   
\begin{tabular}{l}
Diffusion Tensor Imaging (DTI)\\
(weighted least-squares)\\ \citep{Basser1994, Salvador2005}
\end{tabular}  & $\log S(\vecq)$ & $1, q_iq_j$ & $\log S_0, D_{ij}$ & $\text{diag}\left(S^2\right)$ & 0\\[15pt]
\begin{tabular}{l}
Diffusion Kurtosis Imaging (DKI)\\
(weighted least-squares)\\ \citep{Veraart2013}
\end{tabular}
 & $\log S(\vecq)$ & \begin{tabular}{c}
 $1, q_iq_j,$\\
 $q_i q_j q_k q_l$
 \end{tabular}  & \begin{tabular}{c}
 $\log S_0, D_{ij},$\\
 $ K_{ijkl}$ 
\end{tabular} & $\text{diag}\left(S^2\right)$ & 0\\[15pt]
\begin{tabular}{l}
Q-space Trajectory\\
Imaging (QTI)\\ \citep{Westin2016}
\end{tabular} & $\log S(B)$ & $1, B, B^{\otimes 2}$ & $\log S_0, \langle D\rangle,  \mathbb{C}$ & $\text{diag}\left(S^2\right)$ & 0\\[15pt]
\begin{tabular}{l}
Constrained Spherical\\
Deconvolution (CSD)\\ \citep{Tournier2007}
\end{tabular}
& $S(\vecu)$ & $\chi(\vecu)Y_{l}(\vecu)$ & $c_{l}$ & $I$  & L\\[15pt]
\begin{tabular}{l}
Q-Ball Imaging (QBI)\\ \citep{Descoteaux2007, Tuch2004}
\end{tabular} & $S(\vecu)$ & $Y_{l}(\vecu)$ & $c_{l}$ & $I$ & $\int_ {S^2}\|\Delta_b S\|^2 d\Omega$
\\[3pt]
\begin{tabular}{l}
MAP-MRI with Laplacian\\
regularization (MAPL)\\ \citep{Fick2016,Ozarslan2013}
\end{tabular} & $S(\vecq)$ & $\Phi_n(u, q)$ & $c_n$&$I$ & $\int_ {\mathbb{R}^3}\|\Delta S\|^2 d\vecq$\\[15pt]
\begin{tabular}{l}
Spherical Polar Fourier (SPF)\\ \citep{Assemlal2009}
\end{tabular}& $S(\vecq)$ & $R_k(q)Y_{l}(\vecu)$ & $c_{kl}$ & $I$ & $\Lambda_R + \Lambda_Y$ 
\end{tabular}
}
\caption*{
    \begin{tabular}{r l}
      $y$: & response variable\\
      $\vecx$: & input variables\\
      $\phi(\vecx)$: & basis function\\
      $\vecw$: & coefficients\\
      $W$: & inverse noise correlation matrix\\
      $\Lambda$: & regularization matrix\\
      $S$: & signal\\
      $S_0$: & non-diffusion weighted signal\\
      $\chi$: & single fiber response function\\
      $Y_{l}$: & real spherical harmonics, cf. \citep{Descoteaux2007} \\
      $L$: & matrix determined iteratively using a sparsifying heuristic \\
      $\Delta$: & Laplace operator \\
      $\Delta_b$: & Laplace-Beltrami operator\\
      $\Phi_n$: & Hermite functions scaled by a factor $u$\\
      $R_k$: & Gaussian Laguerre polynomials\\
      $\Lambda_R$: & diagonal matrix penalizing higher radial orders\\
      $\Lambda_Y$: & diagonal matrix penalizing higher angular orders
    \end{tabular}
 }
\caption{An assortment of linear models used in dMRI. We have used the notation $\vecq = q\,\vecu$ where applicable. For clarity, some details have been omitted.}
\label{tab:linearModels}
\end{table}

\section{Theory}\label{sec:theory}
To recapitulate, we focus on linear models fitted with linear least-squares. This might sound restrictive at first but --- as can be seen in Table \ref{tab:linearModels} --- it encompasses several of the most widely used models in dMRI. A broad distinction can be made between those derived from a cumulant expansion of the signal \citep{Basser1994, Salvador2005, Veraart2013, Westin2016} and those where the signal is expanded in a, typically orthogonal, basis \citep{Assemlal2009, Caruyer2012, Descoteaux2006, Descoteaux2007, Fick2016, Hess2006, Ozarslan2013, Tournier2007, Tuch2004}.
Differences aside, all these models assume that, in the absence of noise, it would hold that
\begin{equation}
y(\vecx) = \sum_{i=1}^d c_i \phi_i(\vecx), \label{eq:noiseless}
\end{equation}
where $y$ is the response variable that we want to model, $\vecx$ defines a measurement and $\phi_i(\vecx)$ are (possibly nonlinear) functions with corresponding coefficients $c_i$. 
In a typical scenario, we are given $n$ observations $(\vecx_j, y_j)$ and the objective is to determine an estimate $\hat{\vecw}$ such that $\vecy\approx \Phi \hat{\vecw}$, where we have introduced the matrix $\Phi_{ji}=\phi_i(\vecx_j)$. 
Most of the works citepd in Table \ref{tab:linearModels} simply state their estimate in one of these forms:
\begin{equation}
\hat{\vecw} = 
\begin{cases} 
\left(\Phi^T\Phi\right)^{-1}\Phi^T\vecy & \text{(ordinary least-squares)}\\
\left(\Phi^T\noise\Phi\right)^{-1}\Phi^T\noise\vecy & \text{(weighted least-squares)}\\
\left(\Phi^T\Phi + \Lambda\right)^{-1}\Phi^T\vecy& \text{(Tikhonov regularization)}.
\end{cases}
\end{equation}
These are all special cases of the estimate
\begin{equation}
\hat{\vecw} = \left(\Phi^T\noise\Phi + \Lambda\right)^{-1}\Phi^T\noise\vecy,\label{eq:MAP}
\end{equation}
but what is the probabilistic interpretation? In the next section, we show that equation \eqref{eq:MAP} arises naturally as the maximum-a-posteriori solution of a Bayesian linear regression problem.

\subsection{Least-squares and Bayesian linear regression}\label{sec:BayesianLinReg}
Suppose that the observations are given by adding Gaussian noise $\bm{\epsilon}$ to the model in equation \eqref{eq:noiseless},
\begin{equation}
\vecy(\vecx) = \Phi \vecw + \bm{\epsilon},\qquad \bm{\epsilon}\sim \mathcal{N}(0, \sigma^2\noise^{-1}). \label{eq:linearModel}
\end{equation}
We assume that the matrix $\noise$, determining the structure of the noise, is known but not the scale $\sigma^2$. 
It follows that the likelihood --- the probability distribution of the observations given the parameters --- is
\begin{equation}
p(\vecy|\vecw, \vecx) = \mathcal{N}(\Phi \vecw, \sigma^2\noise^{-1}).\label{eq:likelihood}
\end{equation}
As is typical in the Bayesian formalism, we may specify a prior probability density over the coefficients $\vecw$. Suppose that we use 
\begin{equation}
p(\vecw) = \mathcal{N}\left(0, \zeta^2\Lambda^{-1}\right),\label{eq:cprior}
\end{equation}
where $\Lambda$ is known but not the scale $\zeta$. We may assume that the expected value $\mu_\vecw$ is zero without loss of generality, since it is always possible to subtract $\Phi\mu_\vecw$ from the response variable. Bayes' rule then gives the posterior --- the probability distribution of the coefficients given the data --- as follows
\begin{equation}
p(\vecw|\vecy, \vecx) = \frac{p(\vecy|\vecw, \vecx)p(\vecw)}{p(\vecy|\vecx)}.
\end{equation}
Inserting the expressions for the likelihood and prior from above, it is a straightforward exercise in probability theory \citep{Bishop2006} to show that the posterior distribution is again a normal distribution, given by
\begin{equation}
\begin{aligned}
p(\vecw | \vecy, \vecx) &= \mathcal{N}\left(\bm \mu, \Sigma\right),\\
\bm \mu&= Q^{-1}\Phi^T\noise\vecy,\\
\Sigma &= \sigma^2Q^{-1},
\end{aligned}
\label{eq:posterior}
\end{equation}
where $Q = \Phi^T\noise\Phi + (\sigma/\zeta)^2\Lambda$.
We find that by setting $\zeta=\sigma$, the mean of the posterior distribution \eqref{eq:posterior} becomes equal to the least-squares estimate \eqref{eq:MAP}. Thus, by changing perspective and considering the estimate \eqref{eq:MAP} as the mean of the posterior distribution \eqref{eq:posterior}, we get---almost for free---a quantification of the uncertainty in the prediction. The only unknown variable that we need to estimate is the residual variance, $\sigma^2$.

\subsection{Residual variance}
Before deriving an estimate for the residual variance, we first introduce the smoother matrix $H$, which maps responses to predictions, i.e. $\hat{\vecy}= \Phi\hat{\vecw}=H\vecy$ \citep{Hastie2001}. Then, the covariance of the residuals, $\vecr = \vecy-\hat{\vecy}$, can be written
\begin{align}
\text{Cov}[\vecr]&=\text{Cov}[(I-H)\vecy]\\
&=(I-H)\text{Cov}[\vecy](I-H)^T\\
&=(I-H)\sigma^2\noise^{-1}(I-H)^T\\
&=\sigma^2ZZ^T\label{eq:residualCovariance}
\end{align}
where $Z=(I-H)W^{-\frac{1}{2}}$ and $W^{-\frac{1}{2}}$ denotes a matrix square root such as the one given by the Cholesky decomposition.

For an ordinary linear regression model with $n$ observations and $d$ basis functions, the residual variance is estimated as
\begin{equation}
\hat{\sigma}^2 = \frac{\left\|\vecr\right\|_2^2}{\nu},
\end{equation}
where $\nu = n-d$ is referred to as the degrees of freedom. We generalize the estimate of the residual variance to encompass both weighted- and regularized least-squares, by taking the trace of both sides in equation \eqref{eq:residualCovariance}. This gives
 \begin{equation}
\hat{\sigma}^2 = \frac{\left\|\vecy-\hat{\vecy}\right\|_2^2}{\left\|Z\right\|_\text{F}^2}.\label{eq:sigma}
\end{equation}
In other words, the generalized expression for the the degrees of freedom is 
\begin{equation}
\nu=\left\|Z\right\|_\text{F}^2=\text{Tr}\left[(I-H)\noise^{-1}(I-H)^T\right].\label{eq:nu}
\end{equation}

\subsubsection{Residual bootstrap}
In the results, section \ref{sec:simulationResults}, we compare the uncertainty quantification that follows from the Bayesian interpretation in section \ref{sec:BayesianLinReg} with that given by residual bootstrap \citep{Chung2006}.

Residual bootstrap is a form of model-based resampling \citep{Davison1997}, that has been successfully used for dMRI \citep{Berman2008, Chung2006,Haroon2009, Jeurissen2011}. In residual bootstrap, the normalized residuals (after fitting a model) are resampled, and the fitting procedure is repeated for this new draw, after which the parameter value of interest is recorded. The underlying assumption is that the normalized residuals have identical distributions, so that resampling can be done freely among them. For this to hold true, we see from equation \eqref{eq:residualCovariance} that the appropriate normalization is given by
\begin{equation}
\tilde{\vecr}_i = \frac{\vecr_i}{\sqrt{\strut\left(ZZ^T \right)_{ii}}}\label{eq:modifiedResidual}
\end{equation}
If $\Lambda=0$, one can show by direct computation that $ZZ^T=(I-H)\noise^{-1}$. If, further (as assumed e.g. in DTI), the noise is independent but heteroscedastic, then $\noise$ is diagonal but not constant. The expression for the modified residual \eqref{eq:modifiedResidual} then simplifies to \citep{Chung2006, Davison1997}
\begin{equation}
\tilde{\vecr}_i = \frac{\vecr_i}{\sqrt{\left(1-H_{ii} \right)\noise^{-1}_{ii}}}.
\end{equation}
In other words, equation \eqref{eq:modifiedResidual} actually extends the applicability of residual bootstrap, from the ordinary- or weighted least-squares methods previously described in the literature to cases with correlated noise and/or regularization.

\subsection{Uncertainty of the residual variance estimate}
It would be possible to insert the point estimate \eqref{eq:sigma} of the residual variance in equation \eqref{eq:posterior} to get a closed-form expression for the posterior distribution. However, this neglects the uncertainty in the residual variance estimation. We can handle this uncertainty in a Bayesian way, by specifying a distribution over the variance $\sigma^2$. We then determine the posterior distribution for the coefficients by integrating over $p(\sigma^2)$; this procedure is referred to as marginalization.

In \ref{sec:marginalization} we show that there is a particularly convenient prior (inverse-Gamma distribution) for which the posterior is a multivariate $t$-distribution \citep{Kotz2004, Roth2012},
\begin{equation}
p(\vecw | \vecy, \vecx) = t_\nu(\bm\mu, R),\label{eq:tposterior}
\end{equation}
where the mean $\bm \mu$, the degrees of freedom $\nu$ and the correlation matrix $R$ are given by
\begin{equation}
\begin{aligned}
\bm \mu&= Q^{-1}\Phi^T\noise\vecy,\\
\nu &= \left\|Z\right\|_\text{F}^2,\\
R &= \frac{\nu-2}{\nu}\hat{\sigma}^2Q^{-1}.
\end{aligned}
\label{eq:priorHyperparams}
\end{equation}
We emphasize that this replaces equation $\eqref{eq:posterior}$ as the main expression for the uncertainty in the coefficients. Note that the mean, covariance and degrees of freedom are the same as before, but the $t$-distribution has fatter tails than the Gaussian distribution, reflecting the uncertainty in the variance. As the degrees of freedom increases it becomes more similar to a Gaussian distribution; a rule of thumb is that for $\nu\gtrsim 30$ the $t$-distribution can be replaced by a Gaussian.

\subsection{Corollaries}
We can imagine a multitude of uses for the posterior distribution \eqref{eq:tposterior}.  
\begin{itemize}
\item The multivariate $t$-distribution is closed under affine transformations \citep{Roth2012}, which means any property $\bm \theta = A\vecw + \mathbf{b}$ also follows a multivariate $t$-distribution, namely
\begin{equation}
\bm \theta \sim t_\nu(A\bm\mu + \mathbf{b}, ARA^T).\label{eq:linearQuantity}
\end{equation}
This makes it possible to, for instance, compute credible intervals (confidence intervals).
\item The distribution of a property $\theta = f(\vecw)$, nonlinearly dependent on the coefficients, can be determined analytically in some cases, notably if $f$ is invertible. Regardless, it is always possible to estimate the distribution by sampling from the posterior. 
\item If the coefficients are used to condition a probability distribution $p(\theta | \vecw)$, for instance when sampling fiber orientation in probabilistic tractography, one can, at least formally, write down the expression for the posterior predictive distribution:
\begin{equation}
p(\theta | \vecy, \vecx) = \int p(\theta | \vecw) p(\vecw | \vecy, \vecx) d\vecw.
\end{equation}
\end{itemize}

\subsection{Evaluation}
We base most of our quantitative evaluation on the seemingly self-evident statement that ``in $p$ \% of the cases the correct value should be smaller than the $p$th quantile''.
To formalize this statement, recall the definition of the cumulative distribution function of a random variable $X$: $F_X(x) = \text{Pr}(X \leq x)$.
The quantile function $Q$ is, under mild assumptions, the inverse cumulative distribution function, $Q=F^{-1}$. Stated differently, the quantile function specifies, for a given probability $p$, the value $x$ at which the probability of the random variable is less than or equal to the given probability, $\text{Pr}(X \leq x) = p$. 

In our case, given a measurement $\vecy_j$ we get a posterior distribution of the quantity of interest $\theta_j$ following the line of reasoning presented in the previous sections.  For a fixed probability $p$ the corresponding quantile, $q_j = Q_j(p)$, can be estimated. If we know the correct value $\theta^*$, we can observe empirically how often $\theta^* \leq q_i$, and compare this with the theoretically expected probability $p$. This procedure can be repeated for a range of probabilities $p$, after which the comparison can be presented graphically in a so called P-P (probability-probability) plot \citep{Wilk1968}. 

We also use the interquartile range ($IQR$) to summarize the dispersion in a single number. The interquartile range is defined as the difference between the third and the first quartile, i.e. $IQR=Q(0.75) - Q(0.25)$. For a symmetric distribution it would thus correspond to the 50\% confidence interval. One of its main advantages is that it is more robust to outliers than, for instance, the standard deviation.

\section{Results}
We implemented the core algorithmic ideas by extending the free and open source software project Dipy (Diffusion Imaging in Python) \citep{Garyfallidis2014}. The code for our modified version of Dipy is available online\footnote{\url{https://github.com/jsjol/dipy}}. We made a quantitative analysis of the proposed method on simulated data, and a qualitative analysis on in vivo data from the Human Connectome Project \citep{vanEssen2013}.


\subsection{Simulated data}\label{sec:simulationResults}
We simulated data from single- and double-tensor models. For these models, we can calculate the correct values of several quantities analytically. 

In the single-tensor case the mean diffusivity (MD) and fractional anisotropy (FA) are given, respectively, by
\begin{align}
MD &= \frac{\Tr\left(D\right)}{3},\\
FA &= \sqrt{\frac{1}{2}\left(3 - \frac{\Tr\left(D\right)^2}{\Tr\left(D^2\right)}\right)},\label{eq:md}
\end{align}
where $D$ is the diffusion tensor in question. The above, non-standard, expression for FA elucidates its nonlinear dependence on $D$ \citep{Ozarslan2005}. 

In the double-tensor case, the crossing angle is known by design and can be compared with those returned by peak detection algorithms. In addition, there are some scalar indices that can be calculated analytically \citep{Ozarslan2013}, e.g. the return to origin probability (RTOP)
\begin{equation}
RTOP = \frac{1}{S_0}\int_{\mathbb{R}^3} S(\vecq)\,d\vecq.
\end{equation}

\subsubsection{Single tensor model}
Similarly to \citet{Chung2006}, we simulated signals from three different Gaussian distributions with  
axially symmetric diffusion tensors, each having identical mean diffusivity, $MD= 0.7\cdot 10^{-3}$ mm$^2$/s, but different fractional anisotropy, $FA\in\{0.2, 0.5, 0.8\}$.
Assuming a Stejskal-Tanner type measurement \citep{Stejskal1965}, the latent signal was
\begin{equation}
S(\vecq) = S_0\,e^{-4\pi^2 t_d\, \vecq^T D \vecq},
\end{equation}
where $t_d = \Delta - \delta/3$; here $\Delta$ is the mixing time and $\delta$ the pulse duration. We used the same experimental parameters as in the Human Connectome data described in section \ref{sec:invivo}, but with maximum $b$-value ($b=4\pi^2t_dq^2$) equal to $1000$ s/mm$^2$.

\begin{figure}
\centering
\subfloat[Mean diffusivity ($\text{FA}=0.2$)]{\includegraphics[width=0.5\textwidth]{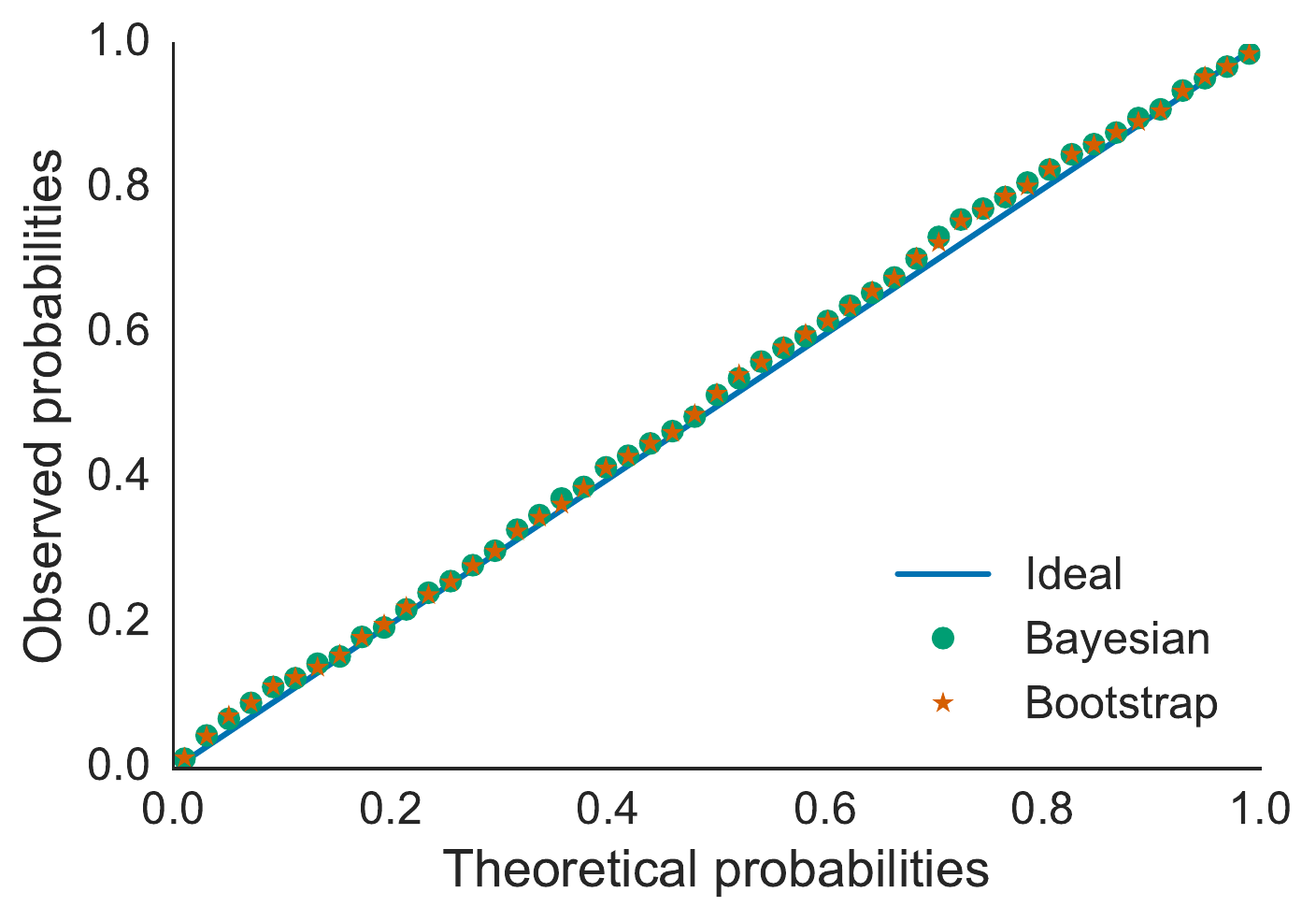}}
\subfloat[Mean diffusivity  ($\text{FA} = 0.5$)]{\includegraphics[width=0.5\textwidth]{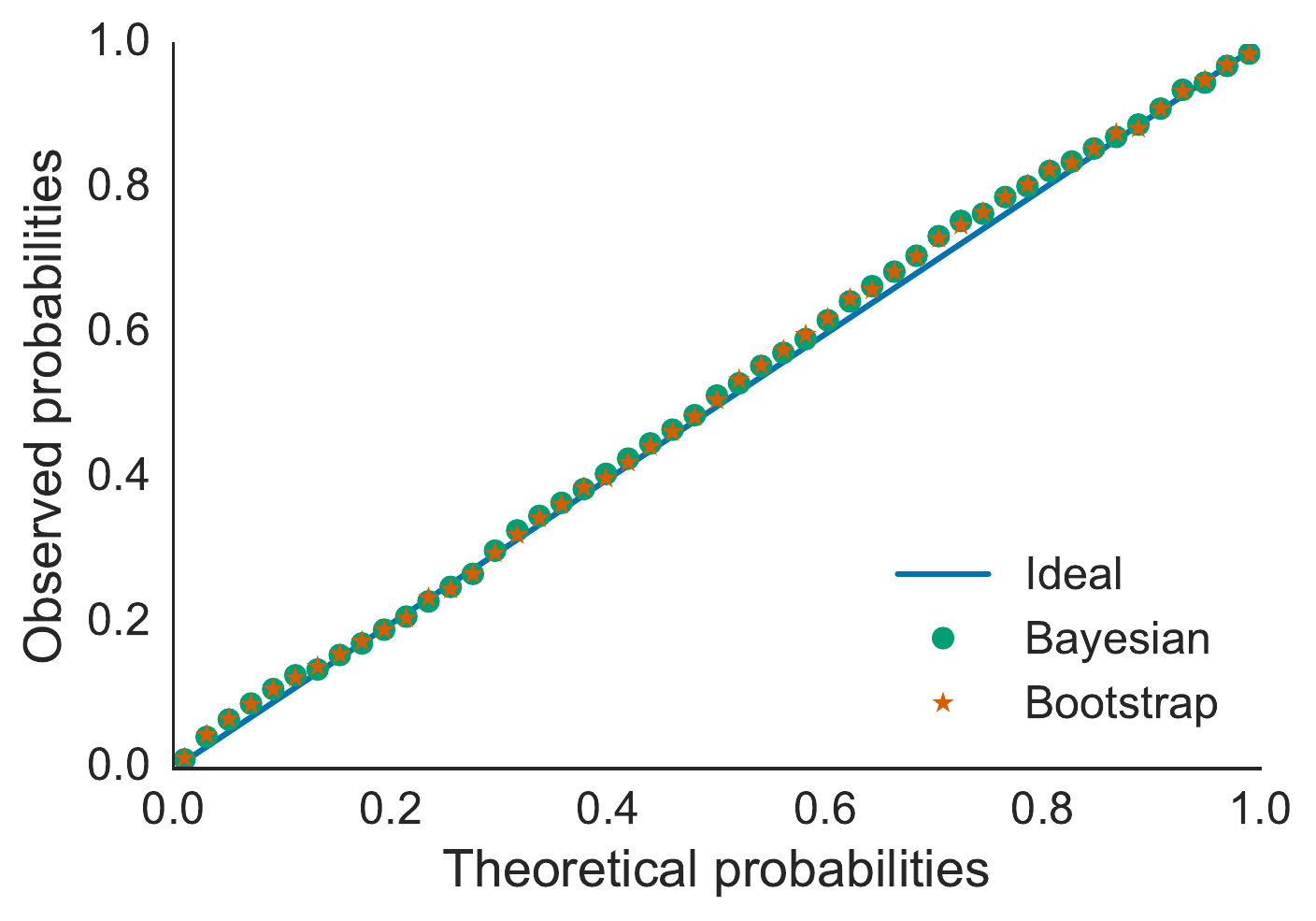}}\\
\subfloat[Mean diffusivity  ($\text{FA} = 0.8$)]{\includegraphics[width=0.5\textwidth]{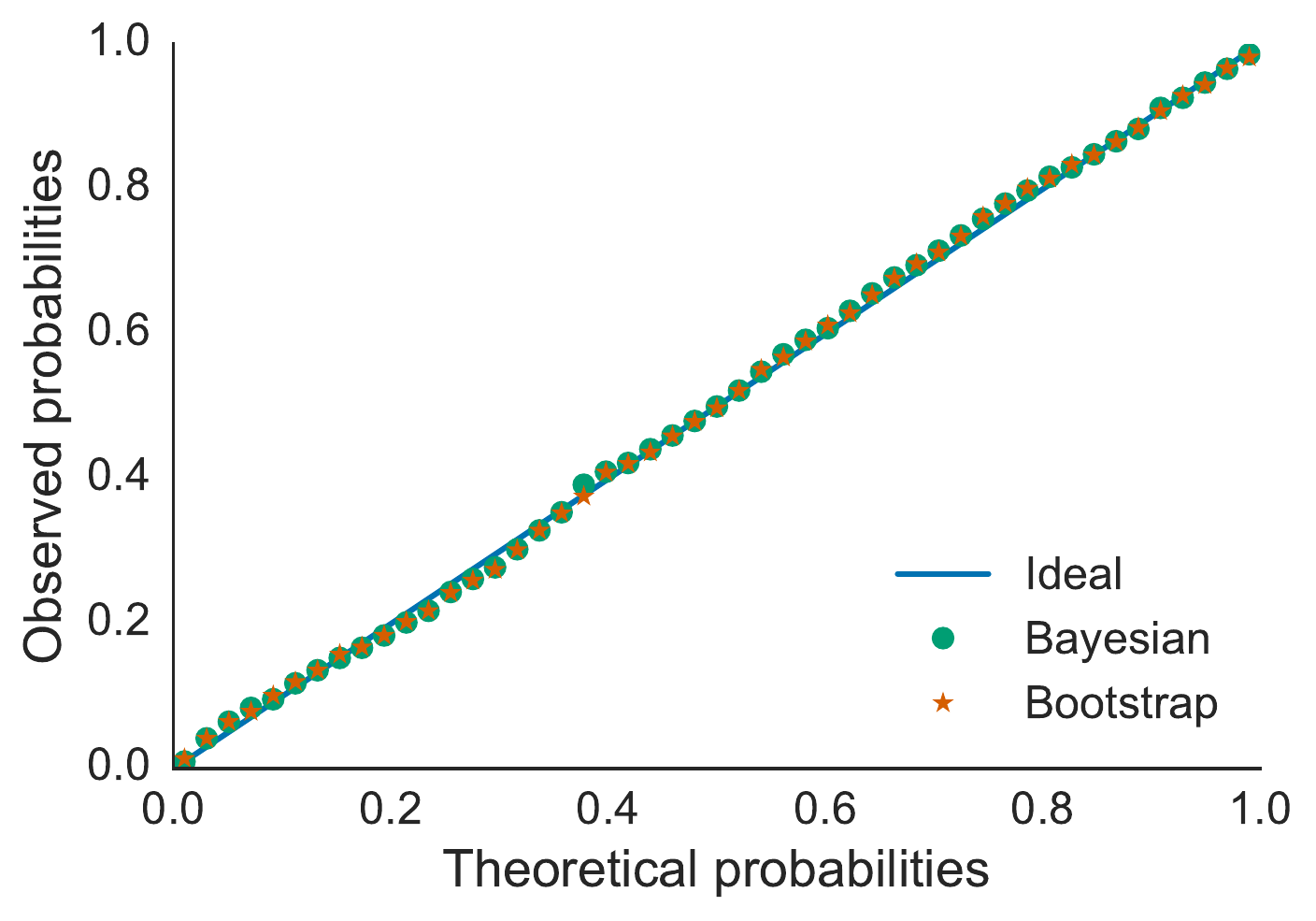}}
\caption{P-P plots of mean diffusivity (MD) for 1000 simulated measurements from three single tensor models with different fractional anisotropy (FA) values. Mean diffusivity was estimated using a DTI model fitted with weighted least-squares. Averaging over the simulations, the P-P plots show how often the correct value is smaller than the quantiles corresponding to the theoretically expected probabilities for that property given a simulated measurement.\label{fig:pp_md}}
\end{figure}

Based on the latent signal, we simulated 1000 independent measurements following a Rician distribution with scale $\sigma = 0.05\, S_0$. For each measurement we used weighted least-squares to fit a diffusion tensor model. The coefficients in the fitting thus include $D_{xx}, D_{yy}$ and $D_{zz}$. Recalling the definition of mean diffusivity \eqref{eq:md}, we see that it is a linear function of the coefficients. Using equation \eqref{eq:linearQuantity} we find that the posterior distribution follows  a univariate $t$-distribution. The P-P plot in Figure~\ref{fig:pp_md} shows the frequency with which the quantiles based on noisy measurements contain the correct value of the mean diffusivity and compares this with 1000 draws using residual bootstrap.

\begin{figure}
\centering
\subfloat[Fractional anisotropy ($\text{FA}=0.2$)]{\includegraphics[width=0.5\textwidth]{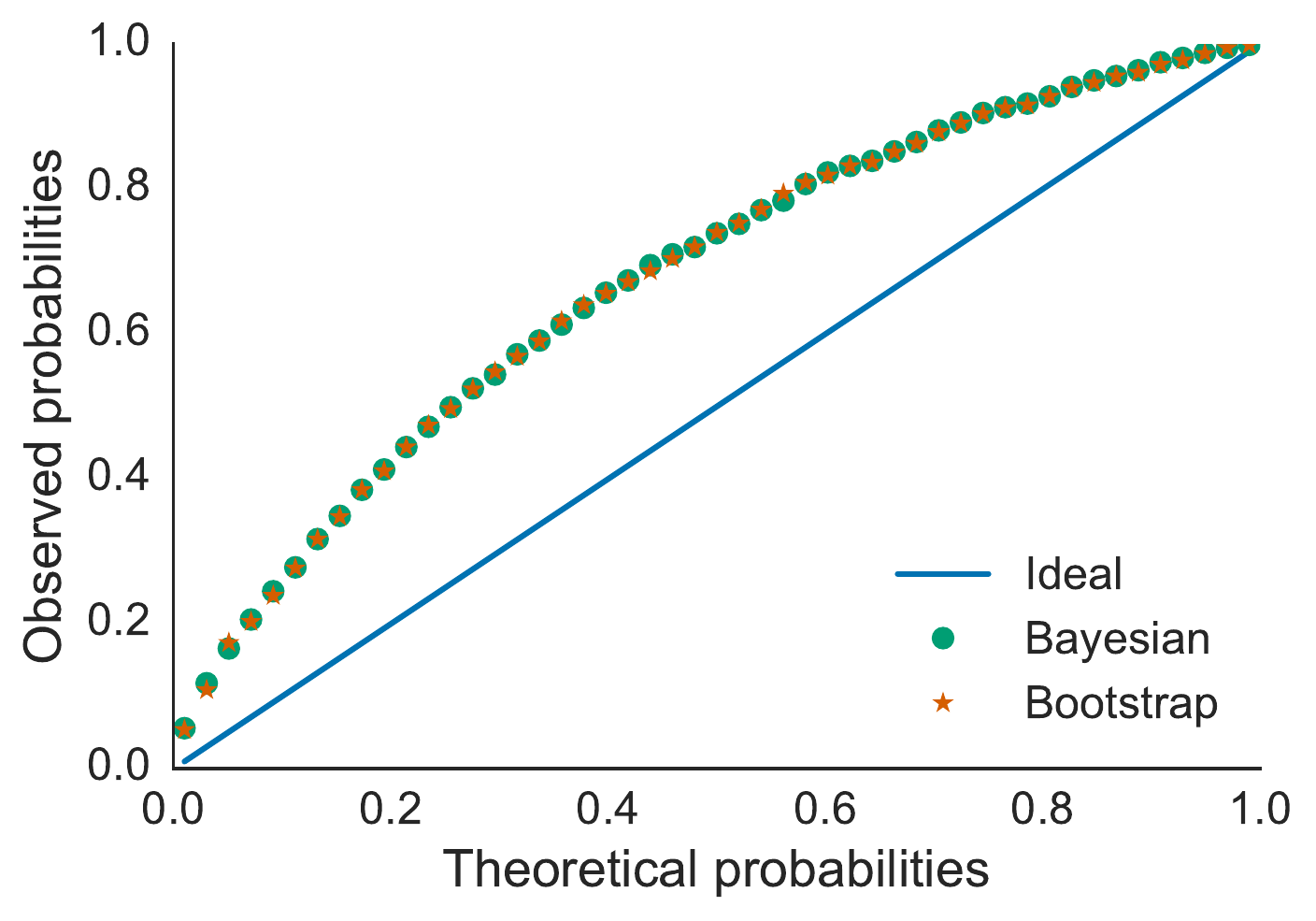}}
\subfloat[Fractional anisotropy ($\text{FA} = 0.5$)]{\includegraphics[width=0.5\textwidth]{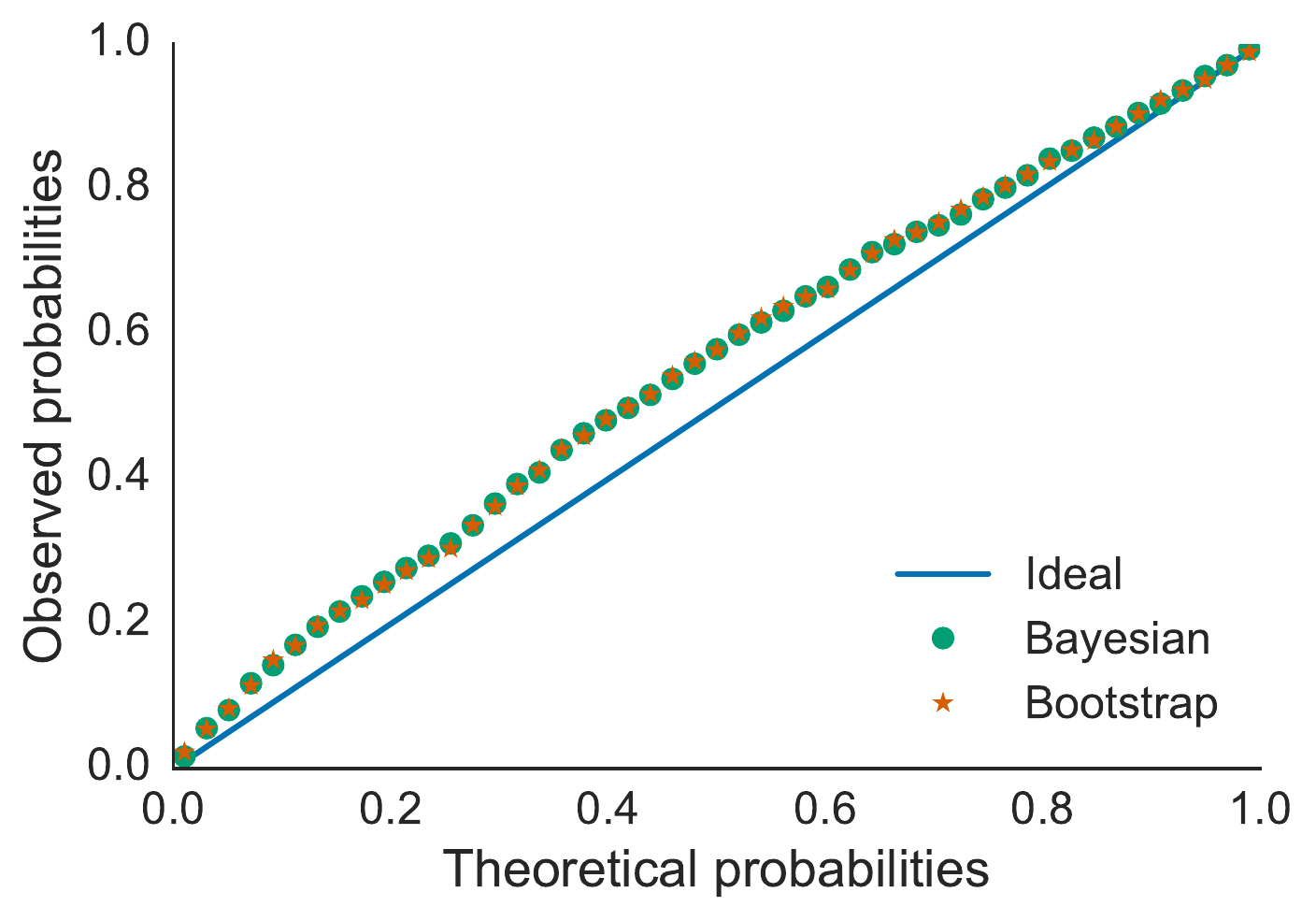}}\\
\subfloat[Fractional anisotropy ($\text{FA} = 0.8$)]{\includegraphics[width=0.5\textwidth]{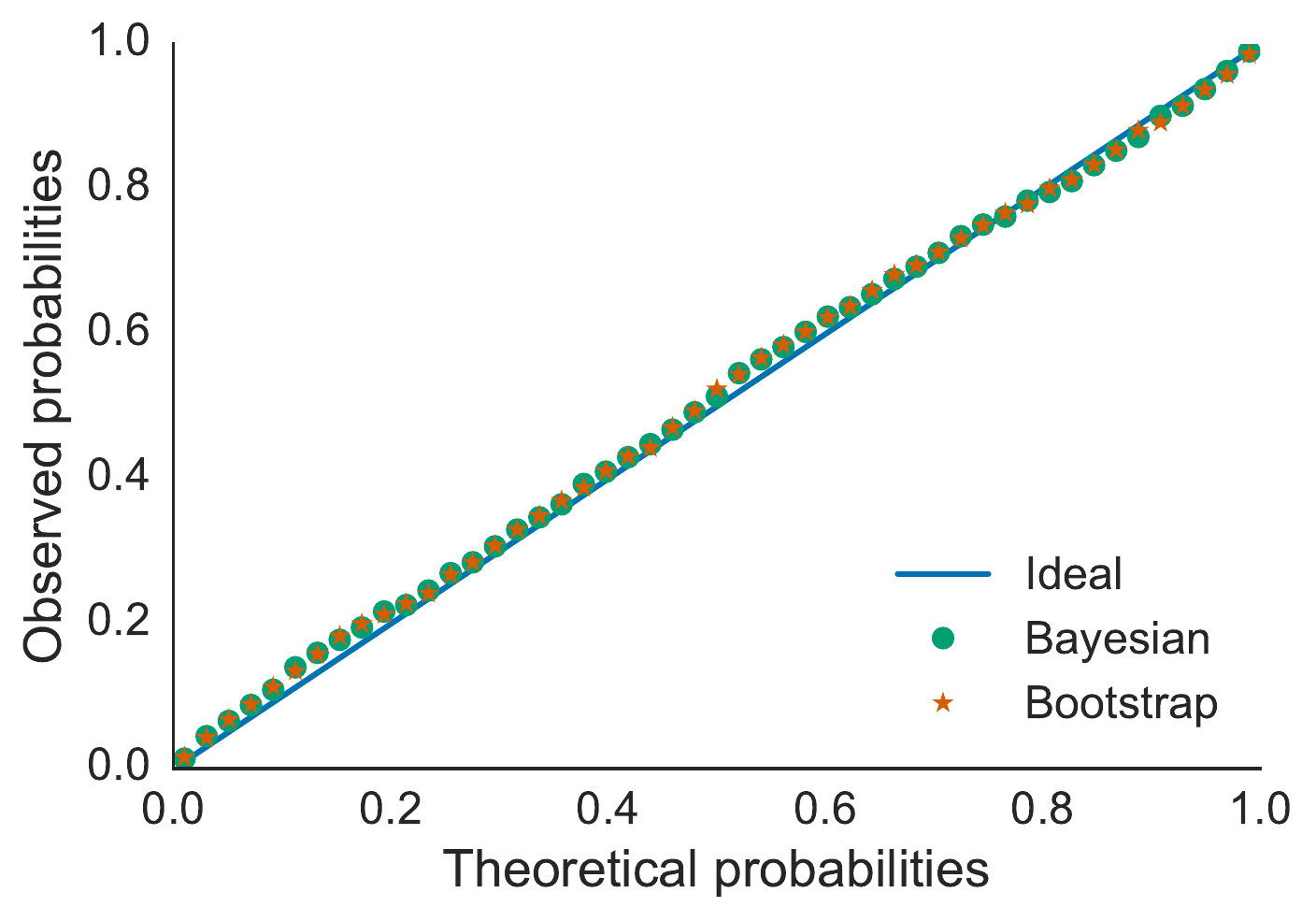}}
\caption{P-P plots of fractional anisotropy (FA) for 1000 simulated measurements from three single tensor models with different FA values. FA was estimated using a DTI model fitted with weighted least-squares. Averaging over the simulations, the P-P plots show how often the correct value is smaller than the quantiles corresponding to the theoretically expected probabilities for that property given a simulated measurement.\label{fig:pp_fa}}
\end{figure}

Fractional anisotropy, on the other hand, is a nonlinear function of the coefficients, which means that we have to resort to sampling to estimate its posterior distribution. For each measurement the quantiles were estimated from 1000 samples from the posterior of the coefficients. The resulting P-P plots are shown in Figure~\ref{fig:pp_fa} and compared with 1000 draws using residual bootstrap.

\subsubsection{Double tensor model}
We defined the double tensor signal as the sum of two, equal magnitude, single tensor signals, but where the second was rotated by either $45^\circ$ or $60^\circ$ about the y-axis. 
The single tensor signal had mean diffusivity $0.7\cdot 10^{-3}$ mm$^2$/s and fractional anisotropy $0.8$, which is roughly equivalent to a white matter tract \citep{Pierpaoli1996}.
Again, we used the same experimental parameters as in the Human Connectome data described in section \ref{sec:invivo}, but this time with maximum $b$-value equal to $3000$ s/mm$^2$. Based on the double tensor signal, we simulated 1000 independent measurements following a Rician distribution with scale $\sigma = 0.05\, S_0$.

The return to origin probability (RTOP) for a multi tensor model with equal diffusion tensors is the same as for a single tensor model, namely
\begin{equation}
RTOP = \text{det}\left(4\pi t_d D\right)^{-1/2},
\end{equation}
which in this case evaluates to $0.90\cdot 10^6$ mm$^{-3}$.

\begin{figure}
\centering
\includegraphics[width=0.8\textwidth]{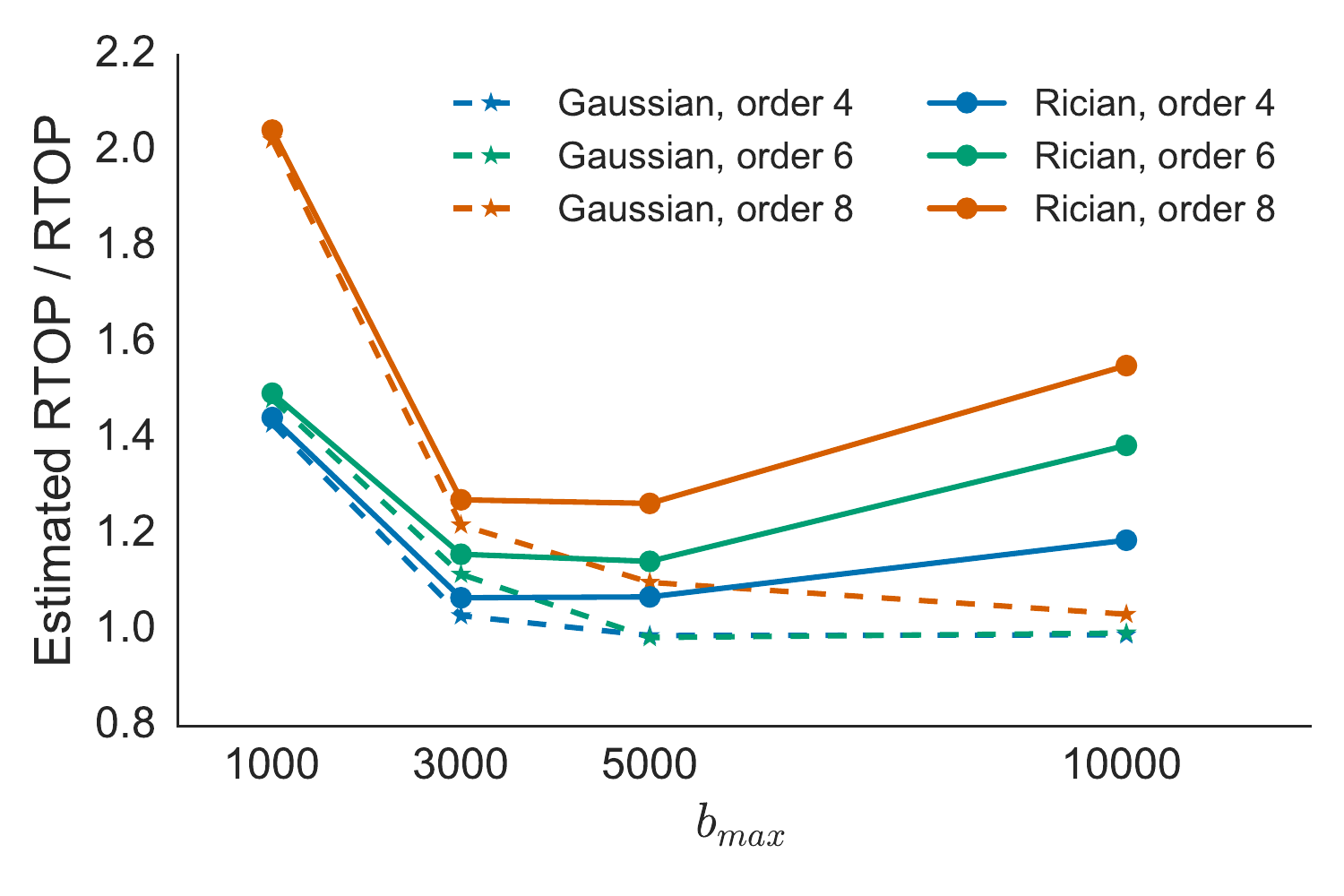}
\caption{Average ratio of estimated to true values of the return to origin probability (RTOP) in 1000 simulated measurements following Gaussian or Rician distributions with scale $\sigma=0.05S_0$.  The signal was generated from a $45^\circ$ double tensor model and was fitted using anisotropic MAPL models of orders 4, 6 and 8 with regularization strength determined by generalized cross-validation.  The experimental parameters were the same as in the Human Connectome Project, except that the maximum $b$-value, $b_\text{max}$, used in the fitting was gradually increased. \label{fig:rtop_bias}}
\end{figure}

\begin{figure}
\subfloat[RTOP ($\theta=45^\circ$)]{\includegraphics[width=0.5\textwidth]{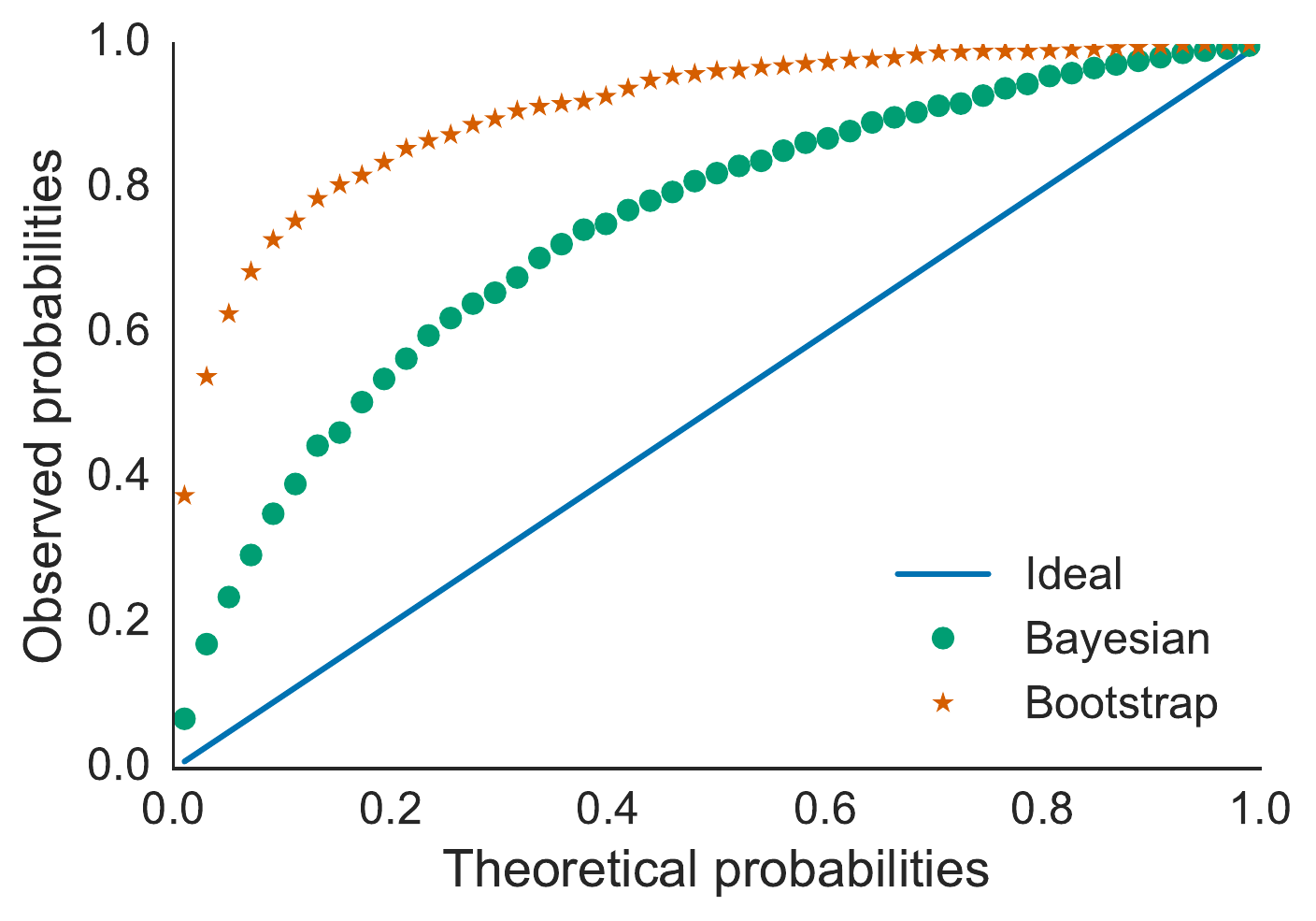}}
\subfloat[RTOP, bias corrected ($\theta=45^\circ$)\label{fig:rtop_bc}]{\includegraphics[width=0.5\textwidth]{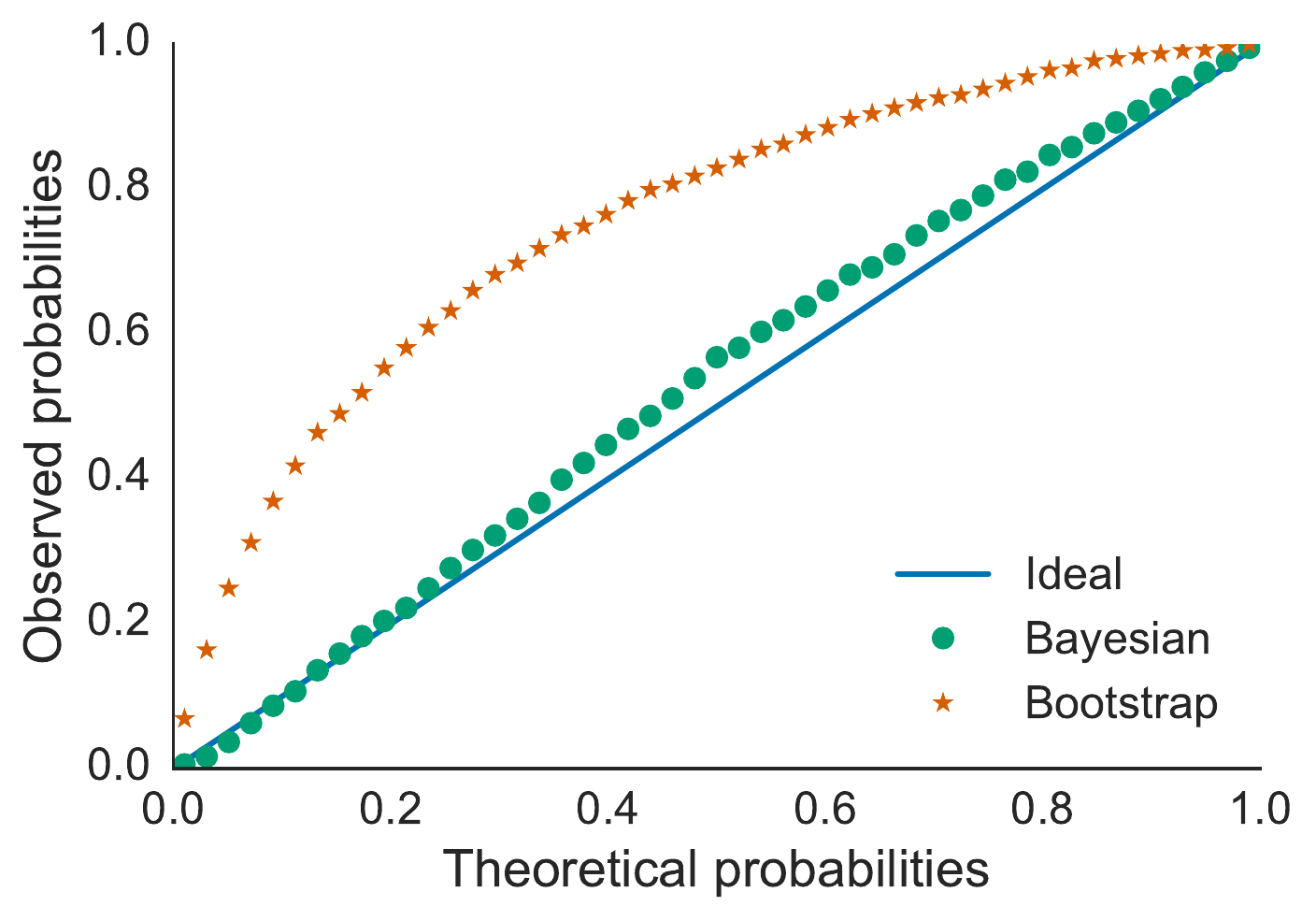}}\\
\subfloat[RTOP ($\theta=60^\circ$)]{\includegraphics[width=0.5\textwidth]{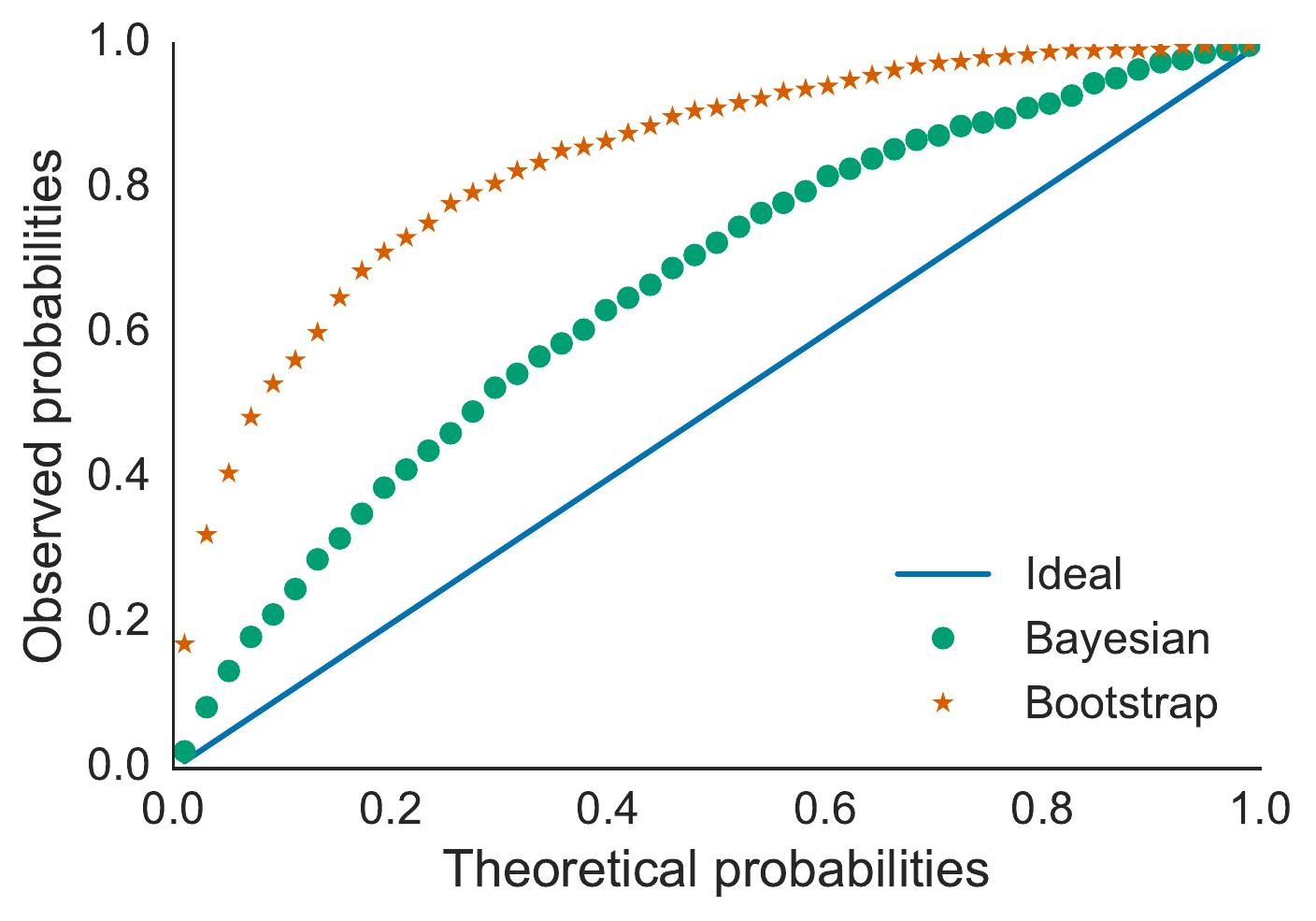}}
\subfloat[RTOP, bias corrected ($\theta=60^\circ$)\label{fig:rtop_bc2}]{\includegraphics[width=0.5\textwidth]{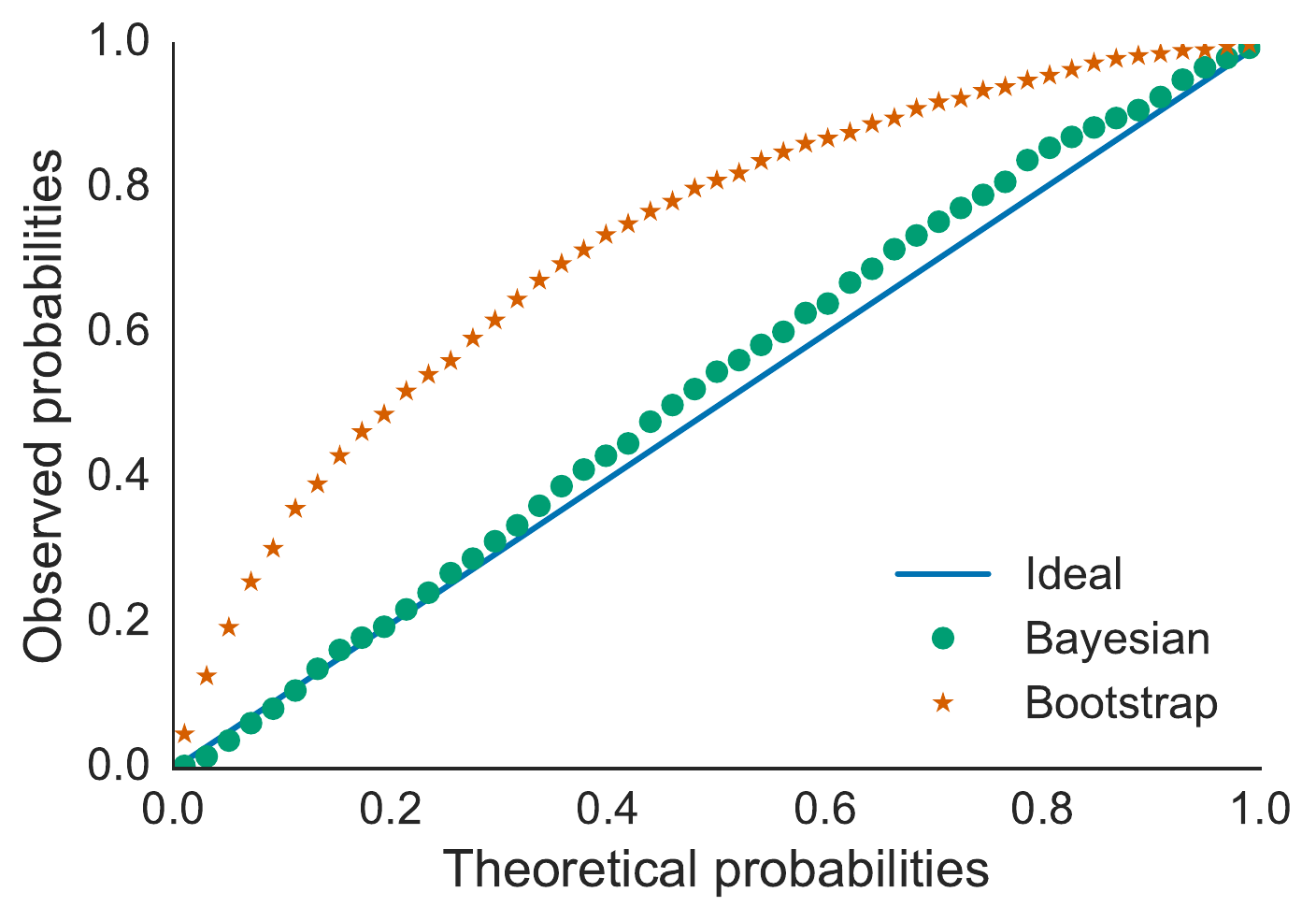}}
\caption{P-P plots of return to origin probability for 1000 simulated measurements from $45^\circ$ and $60^\circ$ double tensor models fitted with anisotropic MAPL models of order 4. Averaging over the simulations, the P-P plots show how often the correct value is smaller than the quantiles corresponding to the theoretically expected probabilities for that property given a simulated measurement. In (b) and (d), the bias (average error of the mean) was subtracted from the theoretical quantiles before comparing with the correct value; this corresponds to shifting the mean of the posterior distribution to the correct value.\label{fig:pp_rtop}}
\end{figure}

For each measurement we fitted a MAP-MRI model \citep{Ozarslan2013} with a Laplacian regularization determined using generalized cross-validation \citep{Fick2016}. For this model the estimate of the RTOP is a linear function of the coefficients, thus equation \eqref{eq:linearQuantity} applies.
We found that the RTOP was on average overestimated. As shown in Figure \ref{fig:rtop_bias}, the size of the bias (difference between average and correct value) depends on both the Rician character of the signal and the maximum $b$-value used. With the settings used in our simulation (in particular, Rician measurements), the bias is smallest when $b\leq 3000$ s/mm$^2$ and the maximum order of the MAP-MRI coefficients is 4. Nevertheless --- even when using these settings --- the bias has an immediate effect on a P-P plot as evident from Figure~\ref{fig:pp_rtop}. We did, however, find that it was possible to remove this confounding effect from the P-P plot by subtracting the bias from the theoretical quantiles before comparing with the correct value, see Figures~\ref{fig:rtop_bc} and \ref{fig:rtop_bc2}. This corresponds to shifting the mean of the posterior distribution to the correct value.
As apparent from Figure~\ref{fig:pp_rtop}, the bias of the 1000 draws using residual bootstrap was larger than that of our Bayesian approach, and the attempted bias correction was not able to compensate for it fully.


\begin{figure}
\subfloat[Crossing angle ($\theta=45^\circ$)]{\includegraphics[width=0.5\textwidth]{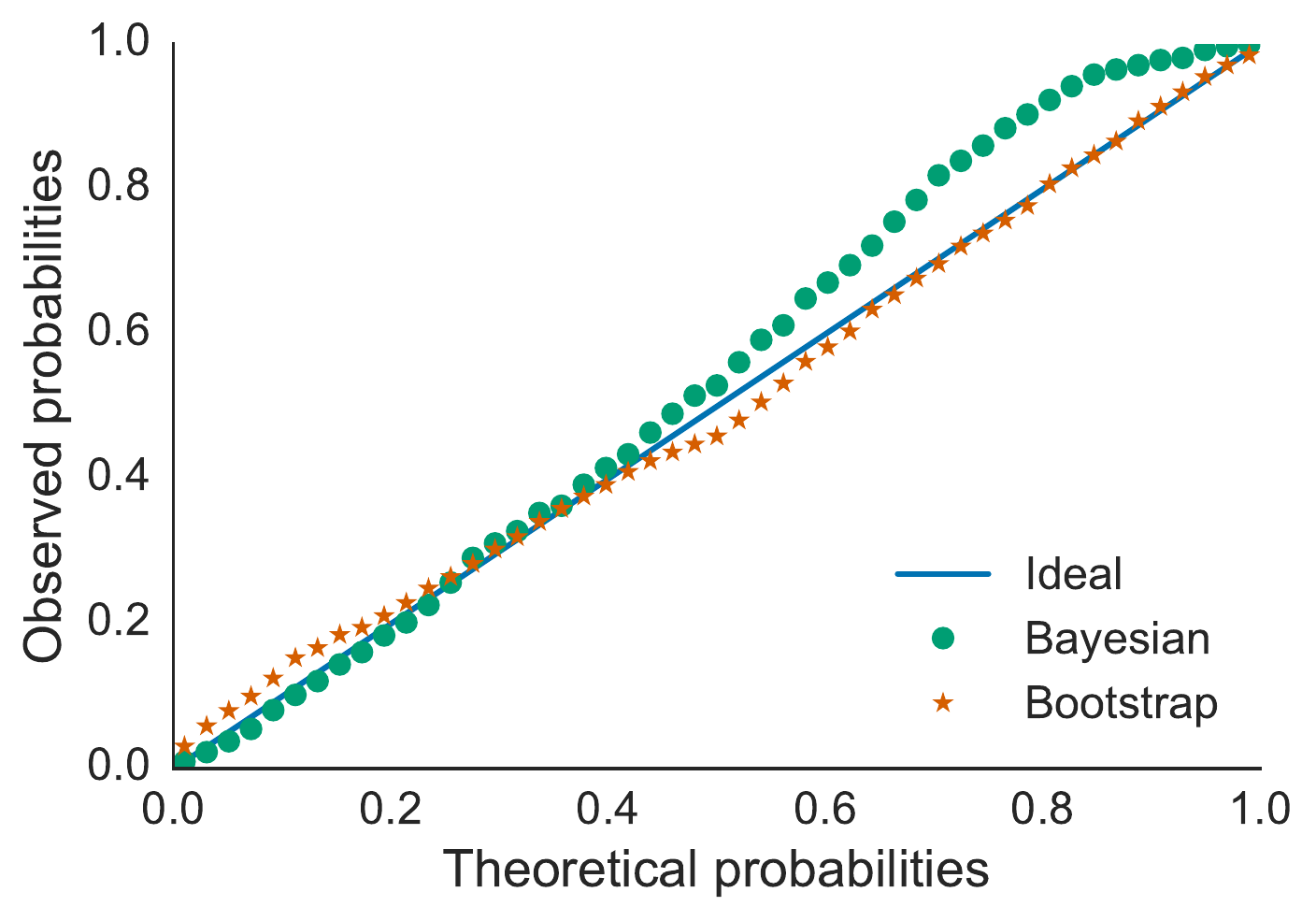}}
\subfloat[Crossing angle ($\theta=60^\circ$)]{\includegraphics[width=0.5\textwidth]{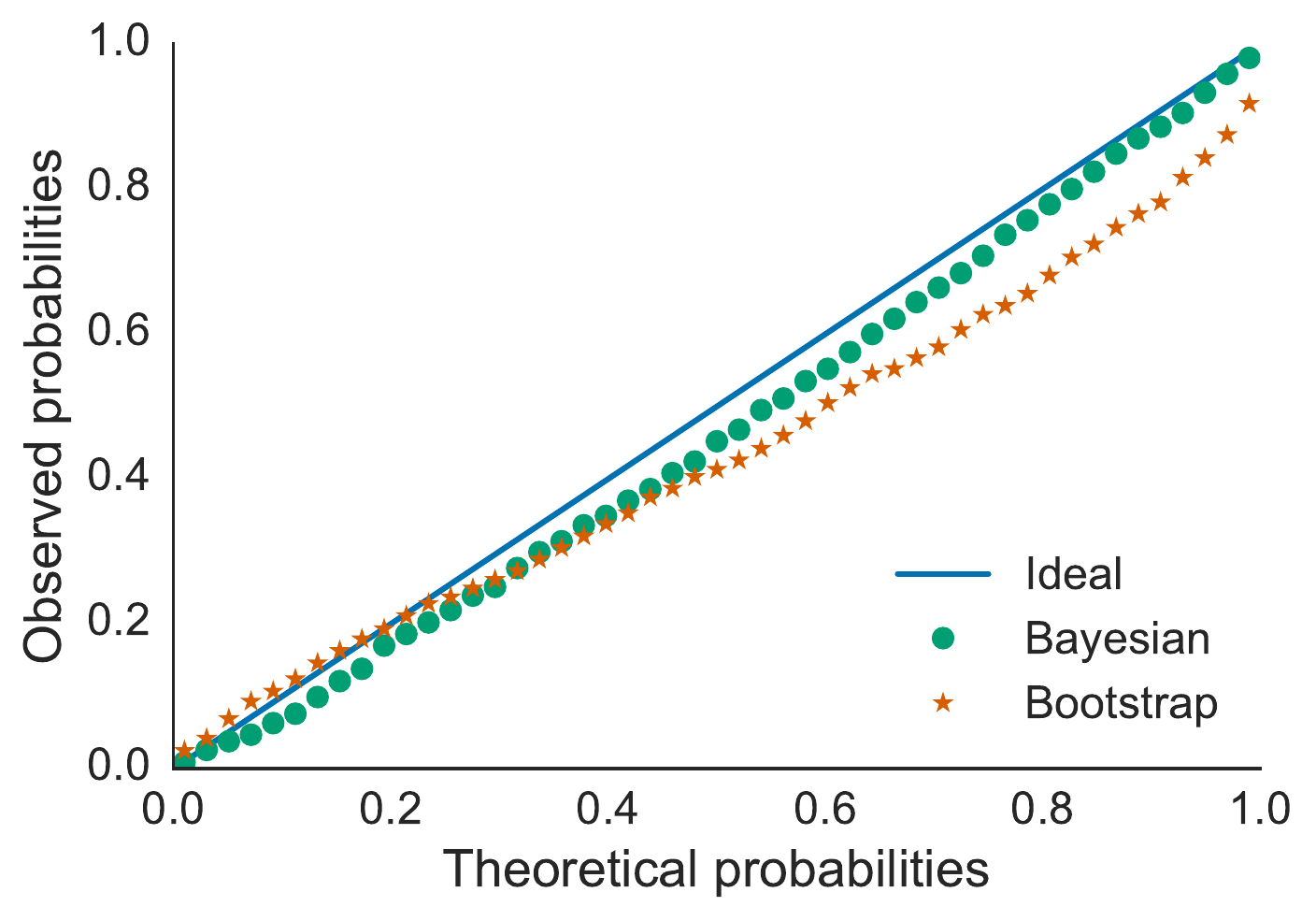}}
\caption{P-P plots of the crossing angle for 1000 simulated measurements from double tensor models fitted with constrained spherical deconvolution (CSD) models of order 10. Averaging over the simulations, the P-P plots show how often the correct value is smaller than the quantiles corresponding to the theoretically expected probabilities for that property given a simulated measurement. 
\label{fig:pp_crossing_angle}}
\end{figure}

Next, we investigated the posterior distribution of the crossing angle when using
Constrained Spherical Deconvolution (CSD) \citep{Tournier2007}. The crossing angle was estimated as the angle between the two largest peaks on the fiber orientation distribution function, provided this angle was at least 25$^\circ$.
We found the peak detection to be more accurate when using only single-shell data. Consequently, we used only the $b=3000$ s/mm$^2$ shell to fit a CSD model of order 10 to each simulated measurement. We set the regularization strength $\lambda$ to 5, and the amplitude threshold $\tau$ (below which the corresponding fiber orientation density is assumed to be zero) to $0.1$. The single tensor model was used as the response function. 
With a crossing angle of $45^\circ$ and $60^\circ$, the peak detection algorithm successfully detected two distinct peaks in 97.4\% and 100\% out of the 1000 cases, respectively. The average estimate of the crossing angle was $47.2^\circ$ and $60.0^\circ$, respectively.
Figure \ref{fig:pp_crossing_angle} shows the resulting P-P plots, and compares with 1000 draws using residual bootstrap.

\subsection{In vivo data}
We show two types of results using in vivo data: first, maps of FA and RTOP with corresponding uncertainties for two healthy adults from the the Human Connectome Project (HCP) \citep{vanEssen2013}; second, the application of our Bayesian approach to a realistic group analysis scenario where we perform voxelwise comparisons of FA between controls and individuals diagnosed with schizophrenia.

\subsubsection{Maps of quantitative features and corresponding uncertainties}\label{sec:invivo}

We used the freely available MGH adult diffusion dataset from the Human Connectome Project\footnote{\url{http://www.humanconnectome.org/documentation/MGH-diffusion/}}\textsuperscript{,}\footnote{\url{https://ida.loni.usc.edu/login.jsp}} (HCP) \citep{vanEssen2013}. The subjects are healthy adults, scanned on a customized Siemens 3T Connectome scanner \citep{Keil2013, Setsompop2013} using a Stejskal-Tanner type diffusion weighted spin-echo sequence. The downloaded data have already been corrected for gradient nonlinearities, subject motion and eddy currents \citep{Andersson2016}. Diffusion measurements were at four b-value shells: 1000, 3000, 5000, 10000 s/mm$^2$. The corresponding number of gradient orientations were 64, 64, 128 and 256. A total of 40 non-diffusion weighted ($b=0$) images were acquired at regular intervals. 

\begin{figure}
\subfloat[Mean FA]{\includegraphics[width=0.5\textwidth]{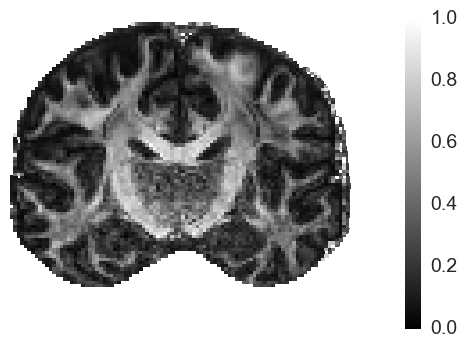}}
\subfloat[Interquartile range]{\includegraphics[width=0.5\textwidth]{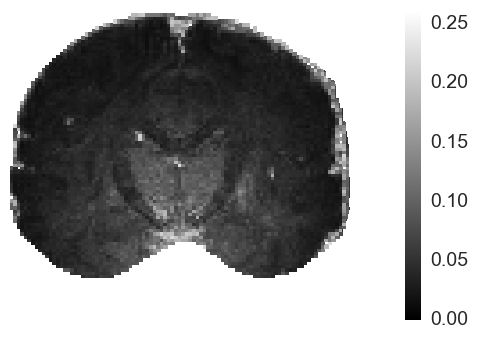}}\\
\subfloat[Mean FA]{\includegraphics[width=0.5\textwidth]{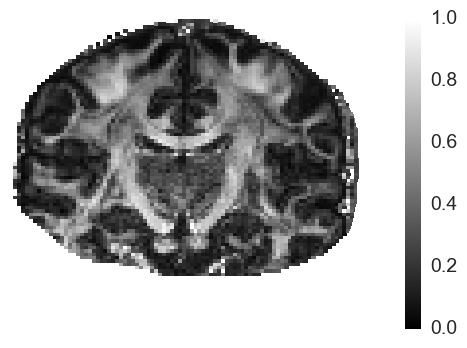}}
\subfloat[Interquartile range]{\includegraphics[width=0.5\textwidth]{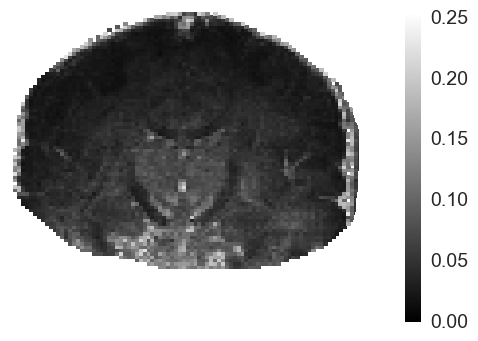}}
\caption{Mean and interquartile range of the fractional anisotropy (FA) estimated using a DTI model fitted with weighted least-squares. The figures correspond to subjects  $\texttt{mgh\_1007}$ (top) and  $\texttt{mgh\_1010}$(bottom) in the MGH adult diffusion dataset, but using only measurements with $b\leq 1000$ s/mm$^2$\label{fig:hcp_fa}}
\end{figure}
\begin{figure}
\subfloat[Mean RTOP]{\includegraphics[width=0.5\textwidth]{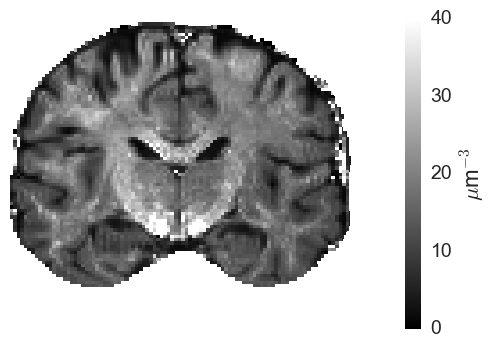}}
\subfloat[Interquartile range]{\includegraphics[width=0.5\textwidth]{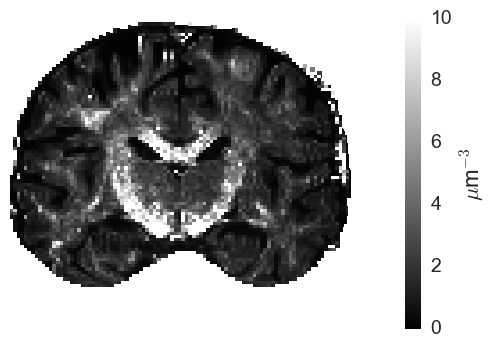}}\\
\subfloat[Mean RTOP]{\includegraphics[width=0.5\textwidth]{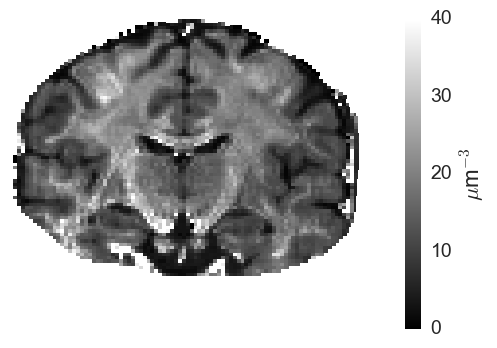}}
\subfloat[Interquartile range]{\includegraphics[width=0.5\textwidth]{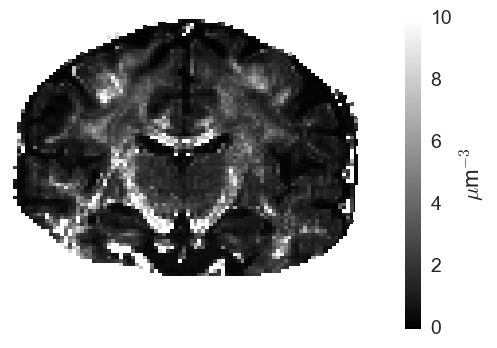}}
\caption{Mean and interquartile range of the return to origin probability (RTOP) estimated using an anisotropic MAP-MRI model of order 6 with Laplacian regularization. The figures correspond to subjects  $\texttt{mgh\_1007}$ (top) and  $\texttt{mgh\_1010}$(bottom) in the MGH adult diffusion dataset, but using only measurements with $b\leq 3000$ s/mm$^2$.\label{fig:hcp_rtop}}
\end{figure}

We analyzed subjects $\texttt{mgh\_1007}$ and $\texttt{mgh\_1010}$. First, we used the same procedure as in the single tensor simulation to fit a diffusion tensor model (weighted least-squares on $b\leq 1000$ s/mm$^2$). We estimated the posterior distribution of fractional anisotropy in each voxel from a draw of 1000 samples.  The mean and the interquartile range of the fractional anisotropy is shown in Figure \ref{fig:hcp_fa}. Second, we used all measurements with $b\leq 3000$ s/mm$^2$ to fit an anisotropic MAP-MRI model of order 6 with Laplacian regularization determined by generalized cross validation. Figure \ref{fig:hcp_rtop} shows the mean and interquartile range of the return to origin probability in each voxel.  On a standard laptop, the computation time needed to estimate FA by fitting a DTI model to each voxel in a slice, such as in Figure \ref{fig:hcp_fa}, was about 3 s whereas sampling in order to estimate the  interquartile range required about 90 s. Similarly, estimating RTOP by fitting a MAP-MRI model required about 220 s whereas obtaining the corresponding, closed-form, estimates of the interquartile range only required 3 s.

\subsubsection{Group analysis of FA differences in schizophrenia}

To apply our Bayesian approach to a realistic group analysis scenario, we downloaded diffusion data from the OpenfMRI project~\citep{openfmri}. Specifically, we downloaded data from the Consortium for Neuropsychiatric Phenomics (CNP)~\citep{Poldrack2016}\footnote{\url{https://openfmri.org/dataset/ds000030/}}, which provides data from 138 healthy controls, 58 individuals diagnosed with schizophrenia, 49 individuals with bipolar disorder and 45 individuals with ADHD. Out of these subjects, we selected 20 healthy controls and 20 individuals with schizophrenia. The CNP dataset contains anatomical $T_1$ volumes, resting state fMRI, task fMRI as well as diffusion MRI. The diffusion data contain 64 diffusion weighted volumes ($b = 1000$ s/mm$^2$) and 1 volume without diffusion weighting ($b=0$), and the voxels have a size of 2 x 2 x 2 mm$^3$.

We corrected the diffusion data for head motion and eddy currents using the function \texttt{eddy\_correct} in FSL (as it is much faster compared to the more advanced eddy function~\citep{Andersson2016}). Our Bayesian approach was then used to generate a posterior distribution of FA for each subject and each voxel inside the brain. The standard TBSS~\citep{Smith2006} pipeline was used to generate a mean FA skeleton, and 1000 draws of FA for each of the 40 subjects were then projected onto this skeleton (by modifying the script \texttt{tbss\_non\_FA}). Two Bayesian group analyses were performed to investigate if there are differences in FA between the two groups. In the first group analysis each subject was treated equally, and each sample in the group posterior $p(\overline{FA}^\text{\,controls} - \overline{FA}^\text{\,schizophrenics} |\, \vecy) $  was calculated using voxelwise, unweighted, means for each group
\begin{equation}
\overline{FA}^\text{\,group}_v = \frac{1}{N^\text{group}}\sum_{i\in\text{group}} FA_{i,v},
\end{equation} 
where  $FA_{i,v}$ represents the posterior distribution for subject $i$ and voxel $v$, and $N^\text{group}$ is the number of subjects in that group. In the second group analysis, we used the uncertainty quantification provided by our Bayesian approach to weight each subject and voxel. We defined the weight $w_{i,v}$ as the reciprocal of the (sample) standard deviation of the corresponding FA posterior distribution~\citep{Beckmann2003}, i.e. as 
\begin{equation}
w_{i,v} = \frac{1}{\text{SD}(FA_{i,v})}.
\end{equation}
A weighted mean was calculated for each group, according to
\begin{equation}
\widetilde{FA}^\text{\,group}_v = \frac{\sum_{i\in \text{group}} FA_{i,v} \cdot w_{i,v}}{\sum_{i\in\text{group}} w_{i,v}}. 
\end{equation}
Then, the differences between the weighted group means were considered as samples from the group posterior.

Finally, a Bayesian t-score (group posterior mean divided by group posterior standard deviation) was calculated in each voxel, since posterior probability maps are rather sensitive to the effect size threshold. For example, calculating the posterior probability $p((\overline{FA}^\text{\,controls} - \overline{FA}^\text{\,schizophrenics}) > 0 |\,\vecy)$ for every voxel often leads to a probability close to 1 in a very large proportion of the voxels, which is not informative. An effect size threshold $\alpha$ can be used to provide more informative posterior probability maps, by instead calculating $p((\overline{FA}^\text{\,controls} - \overline{FA}^\text{\,schizophrenics}) > \alpha |\,\vecy)$, but then the question is what threshold to use. We therefore show maps of Bayesian t-scores, instead of posterior probability maps.

\begin{figure}
\centering
\subfloat[Mean weights healthy controls]{\includegraphics[width=0.5\textwidth]{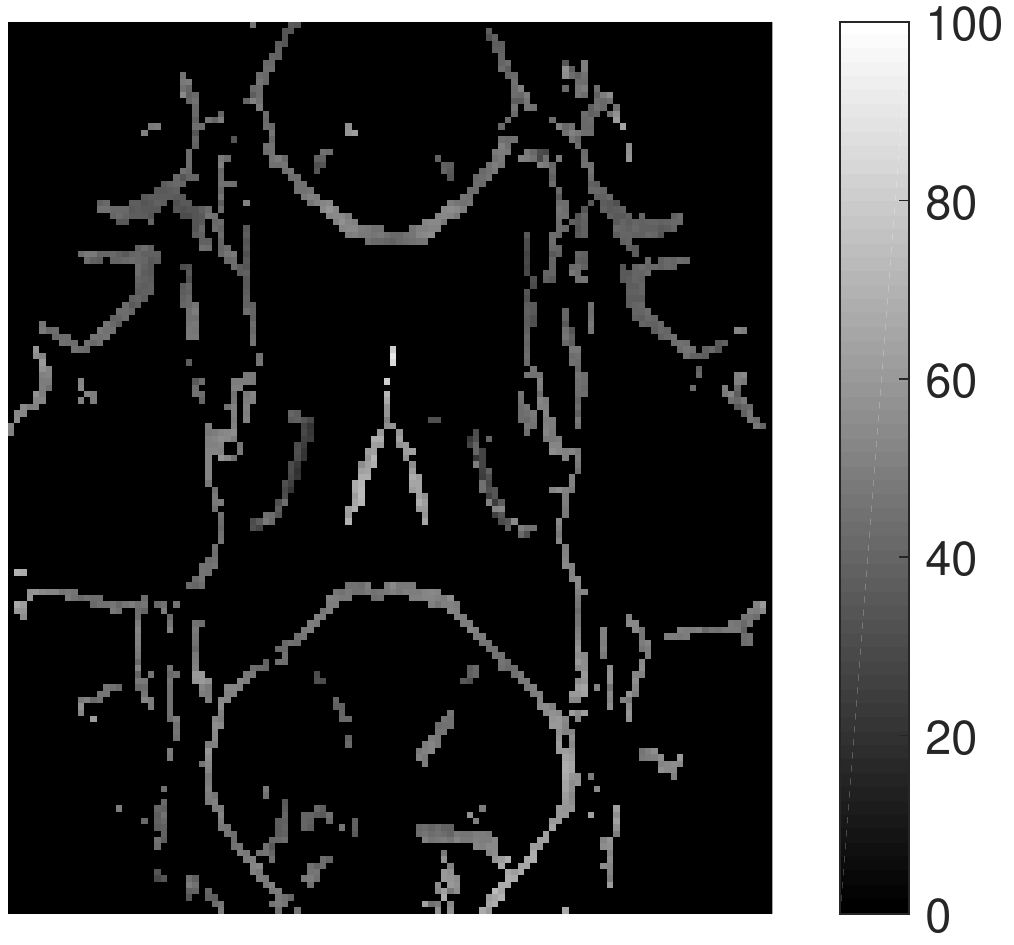}}
\subfloat[Mean weights schizophrenics]{\includegraphics[width=0.5\textwidth]{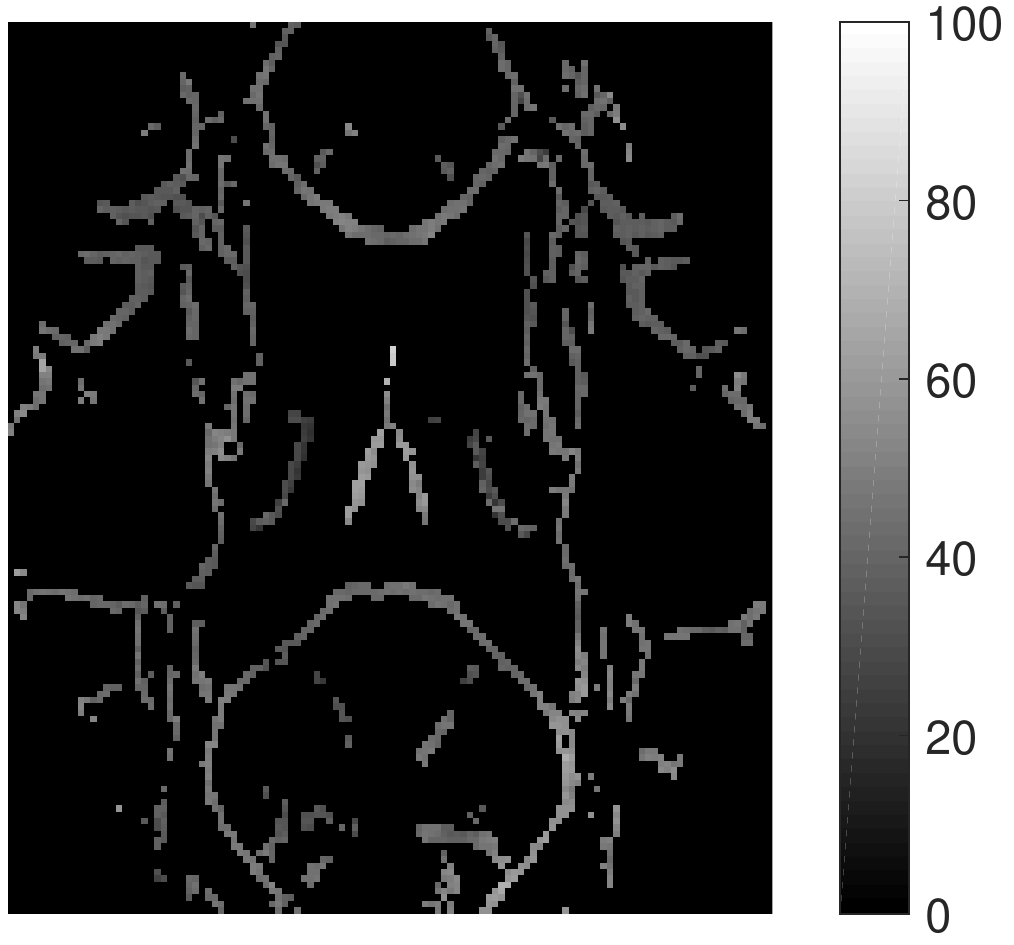}}\\
\subfloat[Controls - Schizophrenics]{\includegraphics[width=0.5\textwidth]{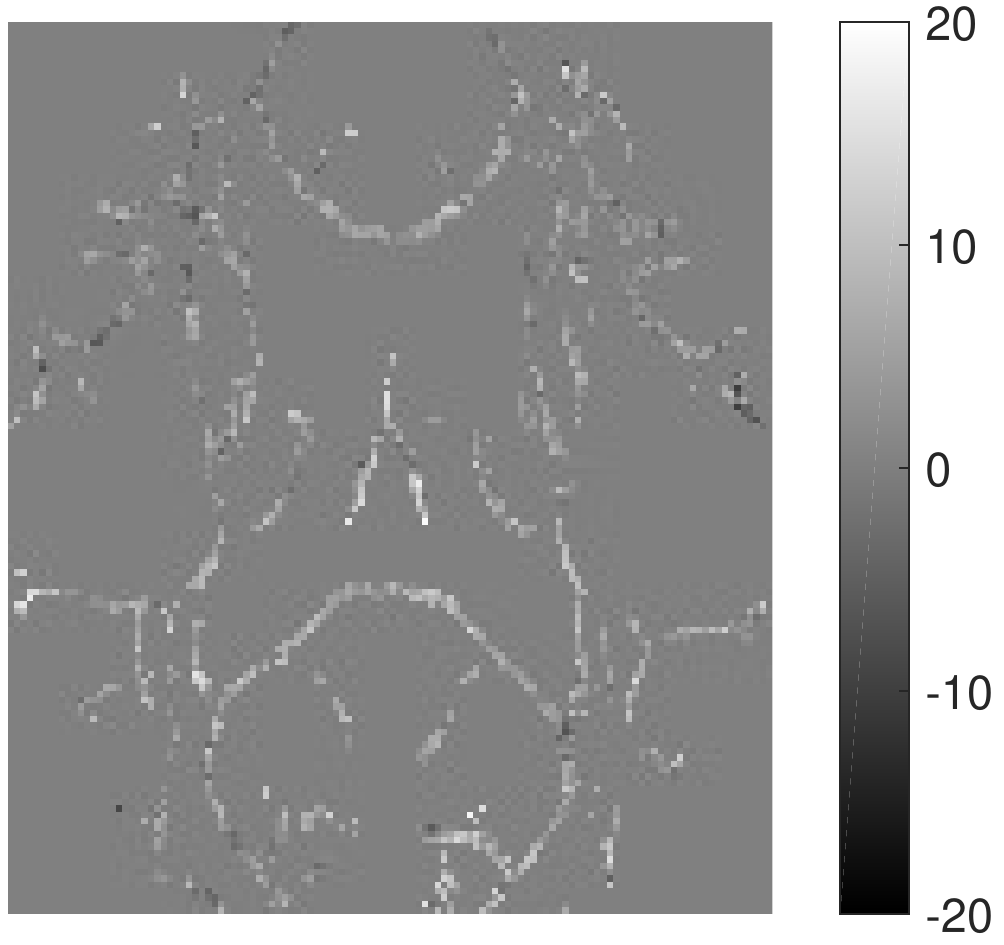}}
\caption{A comparison of weights (calculated as the reciprocal of the posterior standard deviation in each voxel) for healthy controls and individuals diagnosed with schizophrenia. Figures a) and b) show the mean weights over the 20 subjects in each group, while figure c) shows the difference between the mean weights.\label{fig:openfmri_weights}}
\end{figure}

\begin{figure}
\subfloat[Group difference (controls - schizophrenics), unweighted]{\includegraphics[width=0.5\textwidth]{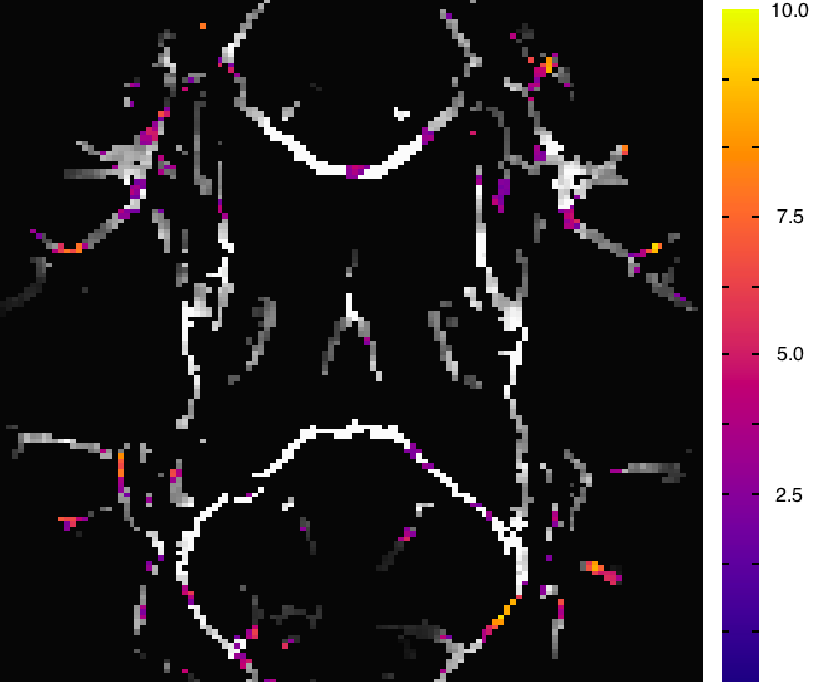}}
 \:
\subfloat[Group difference (controls - schizophrenics), weighted]{\includegraphics[width=0.5\textwidth]{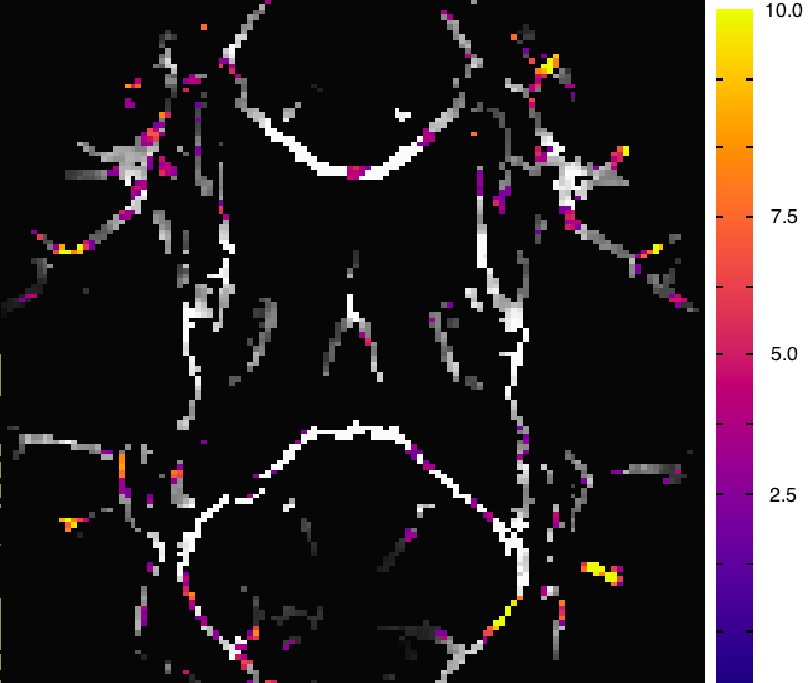}} \\
\begin{center}
\subfloat[Weighted group difference - unweighted group difference]{\includegraphics[width=0.5\textwidth]{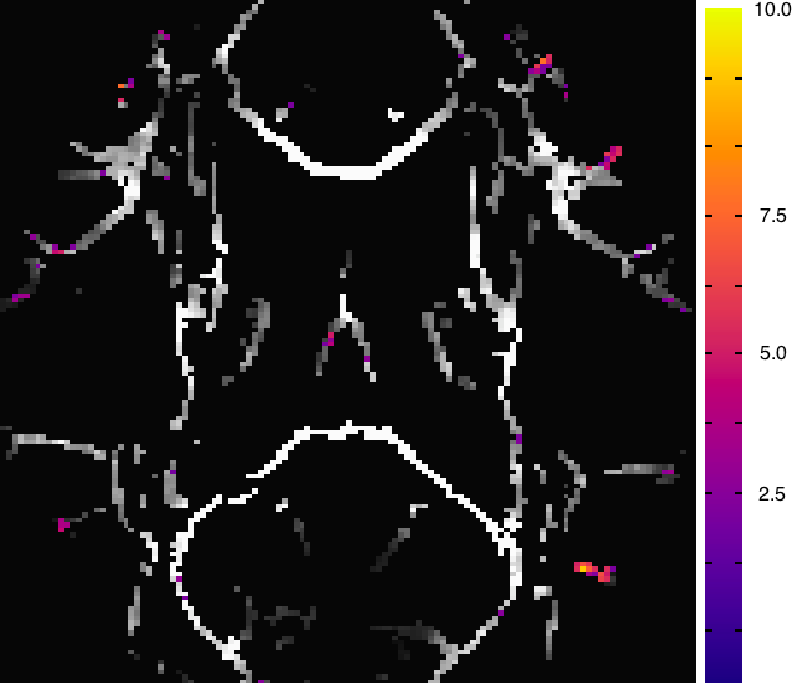}}
\end{center}
\caption{Results from the two Bayesian group analyses; a) unweighted (regular), b) weighted and c) weighted minus unweighted. All maps have arbitrarly been thresholded at a Bayesian t-score of 2.5. The color coding refers to the Bayesian t-score. By downweighting subjects with a high uncertainty, the weighted approach can produce different group results, especially for voxels that are more sensitive to different forms of artefacts (e.g. voxels close to the edge of the brain) and for voxels located in smaller fiber bundles (where FA may be more uncertain). This can be seen in c), as the major differences between the two approaches are located in the outer parts of the skeleton.\label{fig:groupanalysis}}
\end{figure}

\begin{figure}
\centering
\subfloat[FA posterior distributions for all 40 subjects in one voxel (a beta distribution was fitted to the posterior of each subject, using the 1000 draws). One subject with schizophrenia has a much higher uncertainty compared to all other subjects, but this uncertainty is ignored in the unweighted group analysis.]{\includegraphics[width=0.75\textwidth]{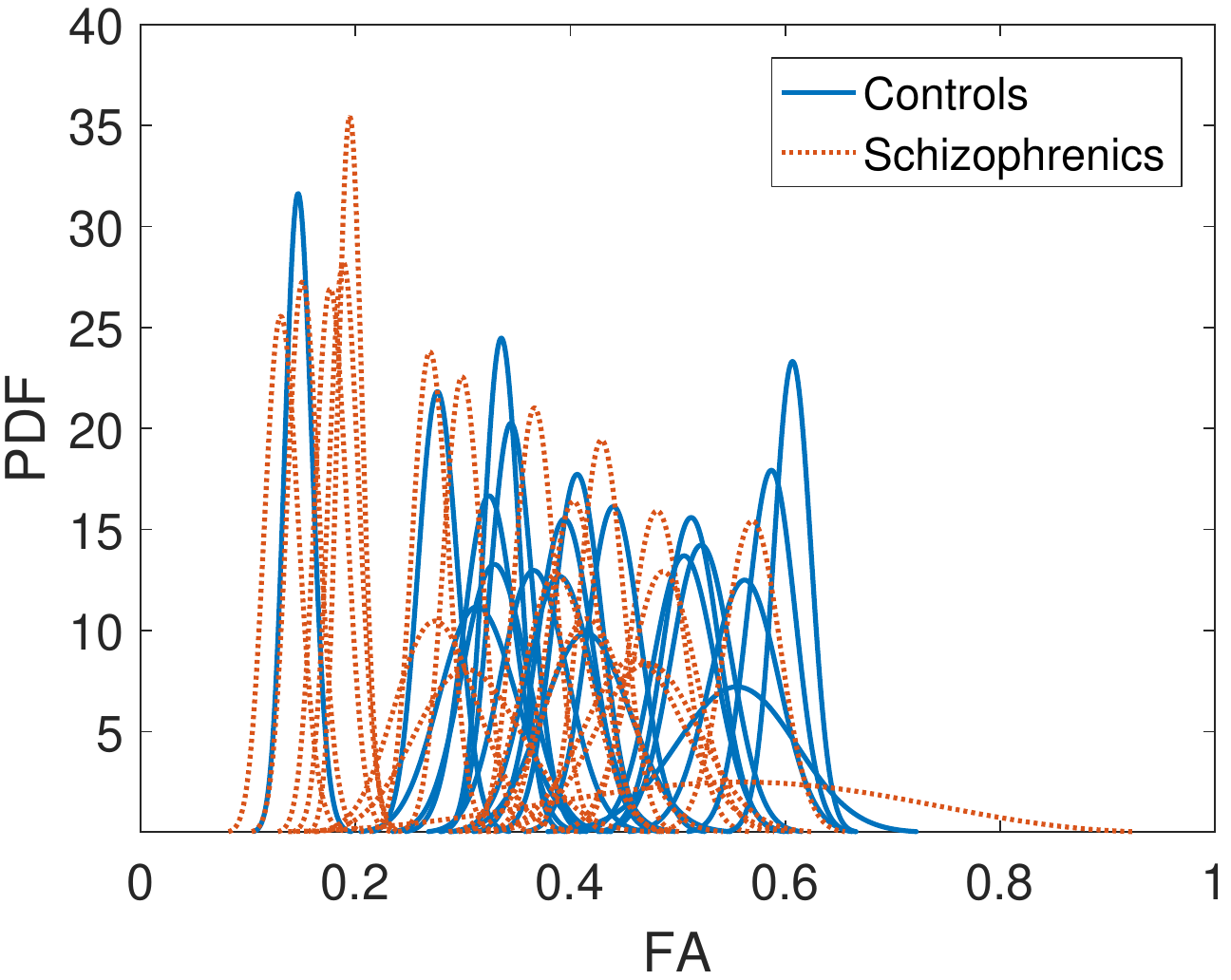}}

\subfloat[Posterior distributions of the group difference in FA (controls - schizophrenics) in one voxel, for the unweighted and the weighted group analysis (a normal distribution was fitted to each posterior, using the 1000 draws).]{\includegraphics[width=0.75\textwidth]{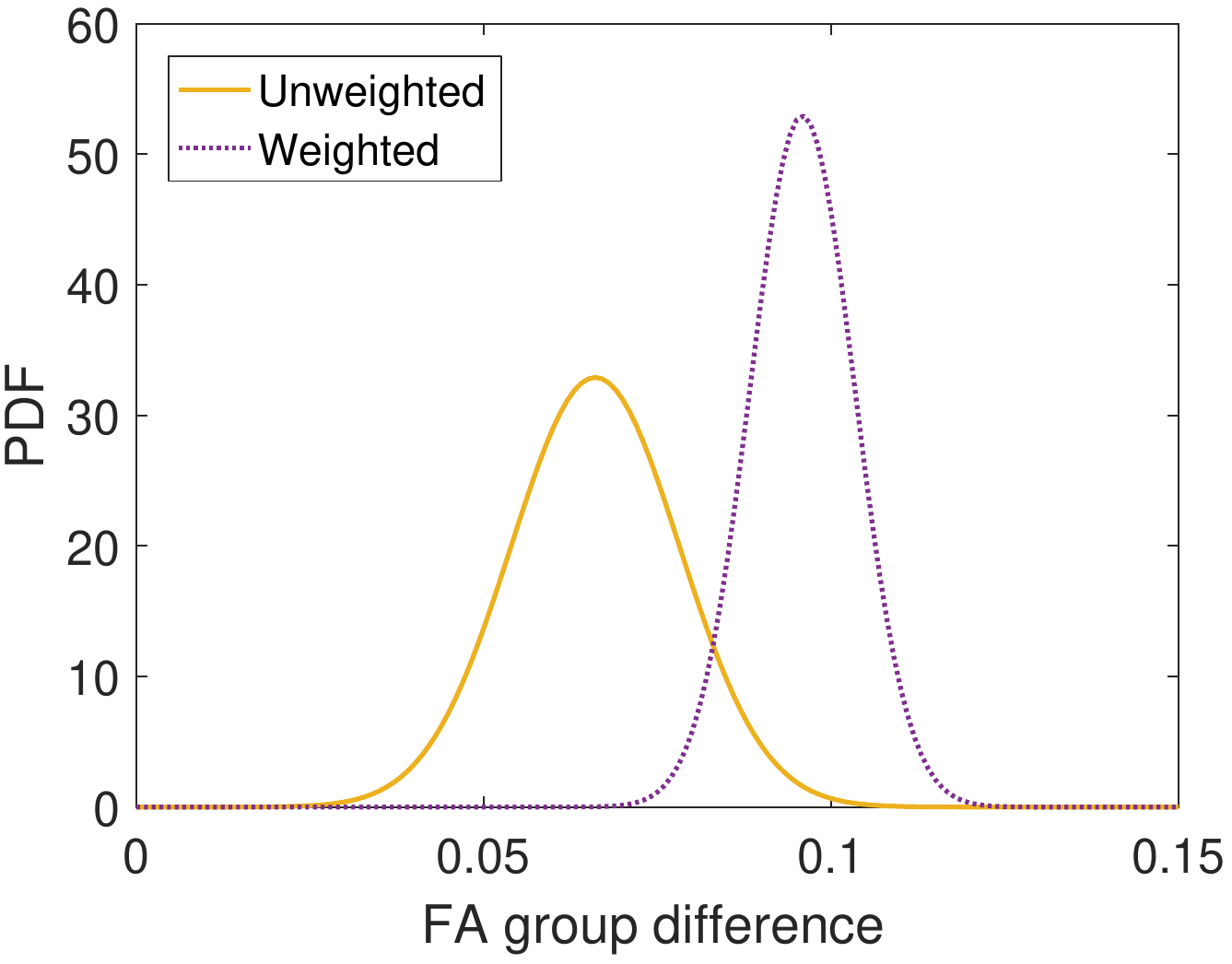}}\

\caption{A comparison of the unweighted and the weighted group analysis in a voxel where the difference between the two approaches is rather large. Subjects with a higher uncertainty will be downweighted in the weighted group analysis, which can lead to a rather different posterior of the group difference. \label{fig:pedagogical}}
\end{figure}

Figure~\ref{fig:openfmri_weights} shows the mean weights for the healthy controls and the individuals diagnosed with schizophrenia, as well as their difference. Figure~\ref{fig:groupanalysis} shows results from the unweighted (regular) as well as the weighted group analysis. Figure~\ref{fig:pedagogical}  shows a (hopefully) pedagogical comparison between the unweighted and the weighted group analysis, for one voxel where the difference between the two approaches is rather large. The weighted group analysis downweights the influence of a subject that has a much higher uncertainty than the others.

\section{Discussion}
\subsection{Interpretation of the results}
For the single tensor simulation fitted with DTI, the P-P plot in Figure~\ref{fig:pp_md}, for the mean diffusivity, shows excellent agreement between theory and observation. This is expected due to the almost perfect match between model and (simulated) measurements --- the signal is generated by a DTI model and for $b=1000$ s/mm$^2$ the signal-to-noise ratio is roughly between 5 and 15, meaning that the likelihood (i.e. measurement distribution) is well approximated as Gaussian \citep{Gudbjartsson1995, Salvador2005}.  
The P-P plot in Figure~\ref{fig:pp_fa}, for the fractional anisotropy, reflects the previously established fact that there is an increasing upward bias of the estimated FA as the true FA decreases \citep{Bastin1998, Farrell2007}. 

From Figures~\ref{fig:pp_md} and \ref{fig:pp_fa} we observe that, in the single tensor simulations, the results of our Bayesian approach and those using residual bootstrap are virtually indistinguishable. The extensive computational requirements of bootstrap methods is one of their main drawbacks. It is thus encouraging that our Bayesian approach offers substantial speed-ups compared to residual bootstrap using 1000 draws---around 20 times in the FA case (that requires sampling) and 200 times in the MD case (using the closed-form expression).

The issue with bias is also present in the RTOP estimation, as can be seen in Figure~\ref{fig:pp_rtop}. After subtracting the bias from the theoretical quantiles, we can see, in Figures~\ref{fig:rtop_bc} and \ref{fig:rtop_bc2}, that there is good agreement between the bias-corrected theoretical distributions and the observed ones. This indicates that the shapes of the posterior distributions are similar to the correct ones, but that their means differ. In other words, the Bayesian uncertainty quantification is accurate although the point estimate (the mean) is not. The results using residual bootstrap, however, exhibit a larger bias which was not fully compensated by the attempted bias correction. We hypothesize that this is because the signal-to-noise ratio differs in the different shells, so that the residuals are not interchangeable (as assumed in residual bootstrap). A potential solution would thus be to only resample within the same shell. There are also more general approaches that could reduce the bias, for instance to denoise the data \citep{Wiest2008} or to transform the signal distribution from Rician to Gaussian \citep{Koay2009}.

In the crossing angle estimation using CSD, Figure~\ref{fig:pp_crossing_angle}, we see that the results of our approach and residual bootstrap are different but of comparable, fairly high, quality. 

For the scalar indices extracted from in vivo data and shown in Figures \ref{fig:hcp_fa} and \ref{fig:hcp_rtop}, we find that the noise levels that can be appreciated qualitatively from the mean images correspond well with larger interquartile ranges. We note that the FA uncertainty appears to increase with the distance from the head's surface. For the RTOP uncertainty, on the other hand, the predominant effect seems to be that the uncertainty correlates with the value of the RTOP. In other words, the RTOP estimates appear to be heteroscedastic.

Although we have not made any effort to optimize our implementation for performance, we argue that the additional computational cost of uncertainty quantification is within reason. In particular, if the model is complex and the quantitiy of interest is an affine function of the coefficients --- as was the case with the RTOP estimation using MAP-MRI --- then the computational overhead of the uncertainty quantification is almost negligible. The FA estimation using DTI, on the other hand, is an example where fitting the model is cheap but estimating the uncertainty requires sampling. 

\subsection{Group analysis}

As can be seen in Figure~\ref{fig:openfmri_weights}, the subject and voxel specific weights differ between the two subject groups. The controls in general have a higher weight (i.e. lower uncertainty). This can for example be explained by differences in the amount of head motion~\citep{Yendiki2014}, or that different MR scanners have different noise levels (at least two MR scanners were used in the CNP study). Existing approaches for group analyses in diffusion MRI, such as TBSS~\citep{Smith2006}, do not use the uncertainty, and can for example not downweight subjects with a high degree of head motion. It is possible that using the more advanced eddy function~\citep{Andersson2016} for preprocessing would lead to more similar weights for all subjects.

Figure~\ref{fig:groupanalysis} indicates minor differences between the unweighted (regular) and the weighted group analysis. The differences are mostly located in the outermost parts of the skeleton, which is consistent with the explanation that they are induced by artefacts. However, as illustrated by Figure~\ref{fig:pedagogical}, a high uncertainty for a single subject can lead to a rather big difference between the two approaches. Both the unweighted and the weighted group analyses indicate a mix of stronger and weaker FA between the groups, while previous schizophrenia studies mainly have found stronger FA in healthy controls~\citep{Kubicki2007}. The standard TBSS group analysis (based on a permutation test) did not reveal any significant group differences, which for example can be explained by low statistical power. The aim of our study was not group differences per se, but to compare the unweighted and the weighted approach.

A possible explanation why weighted group analyses are uncommon in the diffusion MRI field, compared to the fMRI field~\citep{Chen2012,Woolrich2004}, is the higher computational complexity. To generate 1000 draws of FA for one subject took approximately 30 minutes, giving a total of 20 hours for 40 subjects (which is fast compared to Bayesian methods using Markov Chain Monte Carlo (MCMC), often requiring many hours for a single subject~\citep{Wegmann2017}). The TBSS pipeline projects the FA values onto a skeleton where the voxels are 1 x 1 x 1 mm. In our case this resulted in a total of 40 000 volumes with 182 x 218 x 182 voxels each, requiring about 1200 GB of storage. In fMRI the standard is to register each brain to a brain template with an isotropic resolution of 2 or 3 mm, which significantly reduces the memory consumption. Furthermore, it is in fMRI sufficient to store the mean and variance for each voxel (if a normal distribution is assumed), while diffusion MRI may require storing many draws from a more complicated distribution (e.g. FA). The FA distribution in each voxel can be fitted to a beta distribution, as done in Figure~\ref{fig:pedagogical}, which makes it sufficient to store only the two shape parameters in every voxel. Similarly to Bayesian methods for tractography, in FSL based on MCMC~\citep{Behrens2007}, graphics processing units (GPUs) can be used to speedup subject level as well as group level analyses~\citep{Eklund2013,Harms2017,Moises2013}.

\subsection{General reflections}
The class of models fitted with linear least-squares encompasses many of the most popular models. Nevertheless, it would be possible to conduct the same type of probabilistic reasoning for many of the outstanding models, e.g. those using $l_1$-penalization (which would correspond to having a Laplace prior on the coefficients). The catch, however, is that it would no longer be possible to arrive at a closed form expression for the posterior distribution. However, it would still be possible to estimate the posterior by Monte Carlo sampling or approximate methods such as variational Bayes \citep{Gelman2013}.
It could then be viewed as part of a general framework called probabilistic programming \citep{Ghahramani2015}.

In addition to using uncertainty in group analyses and tractography, as we have described before, there are other applications where it could prove useful. One example is in the calculation of the ensemble averaged propagator (EAP),  e.g. as proposed in \citep{Sjolund2017}.


\section{Conclusions}
We make a Bayesian reinterpretation of some of the most popular algorithms for signal estimation in dMRI. In this interpretation, the commonly used point estimate only constitute the mean of a multivariate probability distribution. With access to this distribution it is, in principle, possible to quantify the uncertainty of any derived quantity. In particular, for quantities that are affine functions of the coefficients the posterior can be expressed in closed-form. 
As an example of a dMRI application that we think would benefit in particular from this type of uncertainty quantification, we show results from a realistic group analysis scenario, indicating that a weighted group analysis can indeed downweight subjects with a higher uncertainty.

\section*{Acknowledgements}
This study was supported by the Swedish Foundation for Strategic Research (grant AM13-0090), the Swedish Research Council CADICS Linneaus research environment, the Swedish Research Council (grants 2012-4281, 2013-5229, 2015-05356 and 2016-04482), Link\"oping University Center for Industrial Information Technology (CENIIT), VINNOVA/ITEA3 BENEFIT (grant 2014-00593), and National Institutes of Health (grants P41EB015902, R01MH074794, P41EB015898), and the Knut and Alice Wallenberg Foundation project ``Seeing Organ Function''.

Data collection and sharing for this project was provided by the Human Connectome Project (HCP; Principal Investigators: Bruce Rosen, M.D., Ph.D., Arthur W. Toga, Ph.D., Van J. Weeden, MD). HCP funding was provided by the National Institute of Dental and Craniofacial Research (NIDCR), the National Institute of Mental Health (NIMH), and the National Institute of Neurological Disorders and Stroke (NINDS). HCP data are disseminated by the Laboratory of Neuro Imaging at the University of Southern California.

\section*{Conflict of interest statement}
Declarations of interest: none.

\appendix
\section{Marginalization over the residual variance}
\label{sec:marginalization}
In this section we first show that by placing an inverse-Gamma prior over the residual variance $\sigma^2$, the posterior becomes a multivariate $t$-distribution. Then we show that it is possible to choose the prior hyperparameters such that the covariance and the degrees of freedom remain the same as before putting a prior over $\sigma^2$, ultimately leading to the expressions in equations \eqref{eq:tposterior} and \eqref{eq:priorHyperparams}.

The inverse-Gamma distribution is a convenient choice because it is the conjugate prior for $\sigma^2$. Its probability distribution is
\begin{equation}
p(\sigma^2) = IG(\alpha, \beta) = \frac{\beta^\alpha}{\Gamma(\alpha)}\frac{1}{\left(\sigma^2\right)^{\alpha+1}}\exp\left(-\frac{\beta}{\sigma^2}\right),\label{eq:sigmaPrior}
\end{equation} 
where $\Gamma(\cdot)$ is the Gamma function. 

The joint posterior of $\vecw$ and $\sigma^2$ can be computed by combining the Gaussian likelihood \eqref{eq:likelihood}, the Gaussian prior \eqref{eq:cprior} on the coefficients and the Inverse-Gamma prior \eqref{eq:sigmaPrior} on $\sigma^2$,
\begin{align*}
&p(\vecw, \sigma^2 | \vecy, \vecx) \propto p(\vecy|\vecw, \sigma^2, \vecx)p(\vecw|\sigma^2)p(\sigma^2)\\
& =\frac{|\noise|^{1/2}}{(2\pi\sigma^2)^{n/2}}\exp\left(-\frac{1}{2\sigma^2}\left(y-\Phi\vecw\right)^T\noise\left(y-\Phi\vecw\right)\right)\\
&\cdot\frac{|\Lambda|^{1/2}}{(2\pi\sigma^2)^{d/2}}\exp\left(-\frac{1}{2\sigma^2}\vecw^T\Lambda\vecw\right)
\cdot\frac{{\beta}^{\alpha}}{\Gamma(\alpha)}\frac{1}{(\sigma^2)^{\alpha+1}}\exp\left(-\frac{\beta}{\sigma^2}\right)\\
&\propto (\sigma^2)^{-\left(\alpha+1+\frac{n}{2}+\frac{d}{2}\right)}\exp\left(-\frac{1}{\sigma^2}\left(\beta + \frac{1}{2}\vecy^T W \vecy + \frac{1}{2}\left(\vecw^T Q \vecw - 2\vecw^T\Phi^T\noise\vecy\right)\right)\right)\\
&=(\sigma^2)^{-\left(\alpha+1+\frac{n}{2}+\frac{d}{2}\right)}\exp\left(-\frac{1}{\sigma^2}\left(\beta + \frac{1}{2}\left(\vecy^T W \vecy - \vecy^T\noise\Phi Q^{-1}\Phi^T\noise\vecy\right)\right)\right)\\
&\cdot\exp\left(-\frac{1}{2\sigma^2}\left(\vecw-Q^{-1}\Phi^T\noise\vecy\right)^TQ\left(\vecw-Q^{-1}\Phi^T\noise\vecy\right)\right)\\
&=\mathcal{N}\left(\vecw|\bm \mu, \sigma^2Q^{-1}\right)IG\left(\sigma^2|\alpha_*, \beta_*\right),
\end{align*}
where
\begin{align}
\alpha_*&=\alpha+\frac{n}{2},\label{eq:alphaPosterior}\\
\beta_*&=\beta + \frac{1}{2}\vecy^T\noise\left(\vecy - \Phi \bm\mu\right).\label{eq:betaPosterior}
\end{align}
Now, we obtain the marginal posterior distribution by integrating over $\sigma^2$,
\begin{align}
p(\vecw|\vecy,\vecx)&= \int p(\vecw, \sigma^2 | \vecy, \vecx)\, d(\sigma^2)\\
&=\int \mathcal{N}\left(\vecw | \bm \mu, \Sigma\right)IG\left(\sigma^2 | \alpha_*, \beta_*\right)d(\sigma^2)\\
&=\frac{\Gamma(\alpha_* + \frac{d}{2})}{\Gamma(\alpha_*)}\frac{|Q|^{1/2}}{(2\pi\beta_*)^{d/2}}\left(1+\frac{(\vecw-\bm\mu)^TQ(\vecw-\bm\mu)}{2\beta_*}\right)^{-(\alpha_* + \frac{d}{2})}
\end{align}
which is in fact a multivariate $t$-distribution \citep{Kotz2004, Roth2012} with degrees of freedom $\nu=2\alpha_*$, mean vector $\bm \mu$ and correlation matrix $R = (\beta_*/\alpha_*)Q^{-1}$.
For this result to be usable we must specify the prior hyperparameters $\alpha$ and $\beta$. In the spirit of what is commonly referred to as empirical Bayes, we determine these hyperparameters by requiring that the covariance and the degrees of freedom in equation \eqref{eq:tposterior} remain the same as before putting a distribution over $\sigma^2$.
Setting $\nu=2\alpha_*$, we may write the covariance in \eqref{eq:tposterior} is
\begin{equation}
\frac{\nu}{\nu-2}R=\frac{\beta_*}{\alpha_* - 1}Q^{-1}=E\left[\sigma^2\right]Q^{-1}.
\end{equation}
To get correspondence with the covariance in equation \eqref{eq:posterior}, we require that $E\left[\sigma^2\right] = \hat{\sigma}^2$. Combining this with equations \eqref{eq:betaPosterior} gives
\begin{equation}
\beta=-\frac{1}{2}\left(\frac{1}{\alpha_*}\vecy + \frac{\alpha_*-1}{\alpha_*}\hat{\vecy}\right)^T\noise(\vecy-\hat{\vecy}).
\end{equation}
This means that we can express the degrees of freedom and correlation matrix of the multivariate $t$-distribution in terms of $\nu$ and $\hat{\sigma}^2$ as
\begin{equation}
\begin{aligned}
\nu &= \left\|I-H\right\|_\text{Fro}^2,\\
R &= \frac{\nu-2}{\nu}\hat{\sigma}^2Q^{-1}.
\end{aligned}
\end{equation}

\section*{References}
\bibliographystyle{elsarticle-harv}
\bibliography{refs}
\end{document}